%% file: worrall.tex
\documentclass[multphys,vecphys]{revsvm}

\usepackage{graphicx}    

\begin{document}

\title*{Multiwavelength evidence of the physical processes in radio jets}
\author{D.M. Worrall \& M. Birkinshaw\inst{}}
\institute{HH Wills Physics Laboratory, University of Bristol, Tyndall Ave,
Bristol BS8 1TL, UK
\texttt{d.worrall@bristol.ac.uk, mark.birkinshaw@bris.ac.uk}}
\maketitle

\begin{abstract}
Over the last few years, high-quality X-ray imaging and spectroscopic
data from {\it Chandra\/} and XMM-Newton have added greatly to the
understanding of the physics of radio jets.  Here we
describe the current state of knowledge with an emphasis on the
underlying physics used to interpret multiwavelength data
in terms of physical parameters.
\end{abstract}

\section{Introduction}
\label{sec:introduction}

Jets in active galaxies emit over a wide range of energies from the
radio to the $\gamma$-ray.  Synchrotron radiation and inverse-Compton
scattering are the two main radiation processes, with their relative importance
depending on observing frequency and location within the jet.  The
thermally-emitting medium into which the jets propagate plays a major
role in the properties of the flow and the appearance of the jets.

In this chapter, we concentrate primarily on the interpretation of
observations of resolved jet emission.  Much new information about
jets has been gained over the past few years.
The high spatial resolution of {\it Chandra\/} has been key to the
study of X-ray jets, and the large throughput of XMM-Newton has
assisted studies of the X-ray-emitting environments.  While
historically the search for optical jet emission has been carried out
using ground-based telescopes, a major problem has often been one of
low contrast with light from the host galaxy.  The sharp focus of the
Hubble Space Telescope (HST) helps
to overcome this difficulty, and the HST is playing a particularly
important role in optical polarization studies.  The recent opening of
the Very Large Array (VLA) data archive has greatly assisted
multiwavelength studies of radio jets.  It, together with the
Australian Telescope Compact Array (ATCA) in
the southern hemisphere, allows multifrequency radio mapping at
angular resolutions well matched to {\it Chandra\/} and XMM-Newton.
The combination of data from the radio to the X-ray has been a key
element in advancing our understanding of jets.

Active-galaxy jets provide an exciting laboratory where the interplay
of relativistic effects, plasma physics, and radiation mechanisms can
be seen.  Our purpose in this chapter is to provide an introduction
to the field and some pointers to recent research.

\section{Radiative processes}
\label{sec:processes}

The non-thermal mechanisms of synchrotron radiation and inverse
Compton scattering are described extensively in several places, e.g.,
\cite{blum-gould, ginzburg, longair, pachol, rlightman}, and
descriptions of thermal radiation can be found, e.g., in
\cite{longair, rlightman, sarazin, spitzer}. In the following sections
we give equations important in the interpretation of jet emission in
useful forms that are normally independent of the system of units.

\subsection{Synchrotron radiation}
\label{sec:synchrotron}

The rate of loss of energy of an electron (or positron) of energy $E$
is given by

\begin{equation}
-dE/dt = 2 \sigma_{\rm T} c \gamma^2 \beta^2 u_{\rm B} \sin^2\alpha
\label{eq:synlosspitch}
\end{equation}

\noindent
where $\sigma_{\rm T}$ is the Thomson cross section, $c$ is the speed
of light, $\gamma$ is the Lorentz factor of the electron ($= E/m_{\rm
e}c^2$), $\beta$ is the speed of the electron in units of $c$, $u_{\rm
B}$ is the energy density in the magnetic field, and $\alpha$ is the
pitch angle between the direction of motion and the magnetic field.
Averaging over isotropic pitch angles, $P(\alpha) d\alpha =
(1/2)\sin\alpha\,d\alpha$,

\begin{equation}
-dE/dt = (4/3) \sigma_{\rm T} c \gamma^2 \beta^2 u_{\rm B} 
\label{eq:synloss}
\end{equation}

The radiative lifetime of the electrons is usually calculated as

\begin{equation}
\tau_{rad} = E/(-dE/dt)
\label{eq:tausyn}
\end{equation}

\noindent
which, for energy losses proportional to energy squared, as here,
is the time for any given electron to lose half of its energy.
High-energy electrons, responsible for high-energy radiation, lose
their energy fastest.

The spectral distribution function of synchrotron radiation emitted by
monoenergetic electrons of Lorentz factor $\gamma$ is rather broad.
It is usual to define the critical frequency,

\begin{equation}
\nu_{\rm c} = (3/2) \gamma^2 \nu_{\rm g} \sin\alpha
\label{eq:synfreq}
\end{equation}

\noindent
where $\nu_{\rm g}$ is the non-relativistic electron gyrofrequency,
which is proportional to the magnetic field strength, $B$.  Written in
SI units, $\nu_{\rm g} = eB/2\pi m_{\rm e} \approx 30 B$~GHz, where
$B$ is in units of Tesla.  As a rough approximation, something close
to $\nu_{\rm c}$ can be used as the frequency of emission, but for the
full distribution function it is convenient to define $X =
\nu/\nu_{\rm c}$, and then the spectral distribution function depends
on frequency through

\begin{equation}
F(\nu, \nu_{\rm c}) = X \int_X^\infty K_{5/3}(\zeta) d\zeta
\label{eq:syndist}
\end{equation}

\begin{figure}[t]
\centering
\includegraphics[height=5cm]{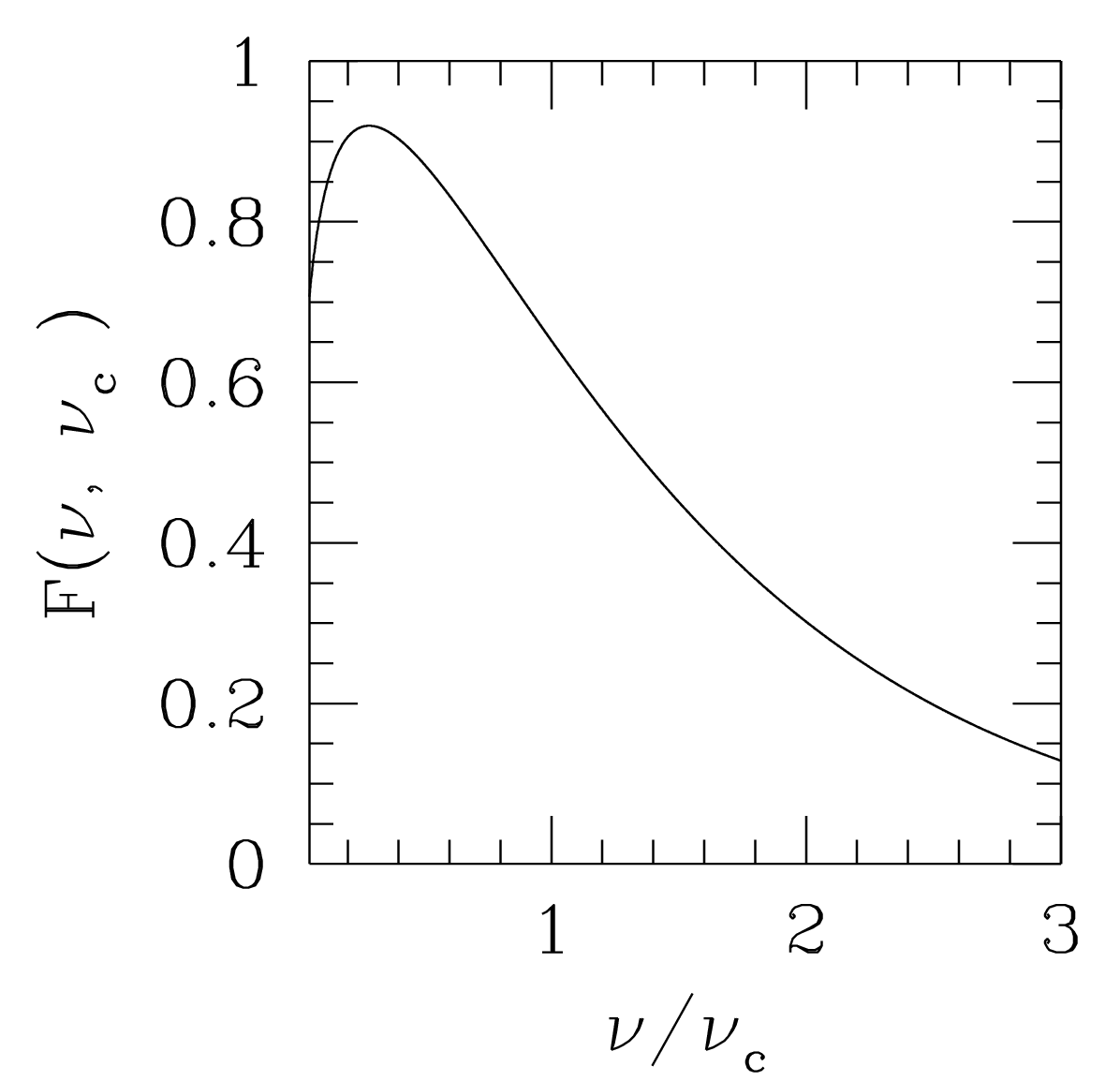}
\includegraphics[height=5cm]{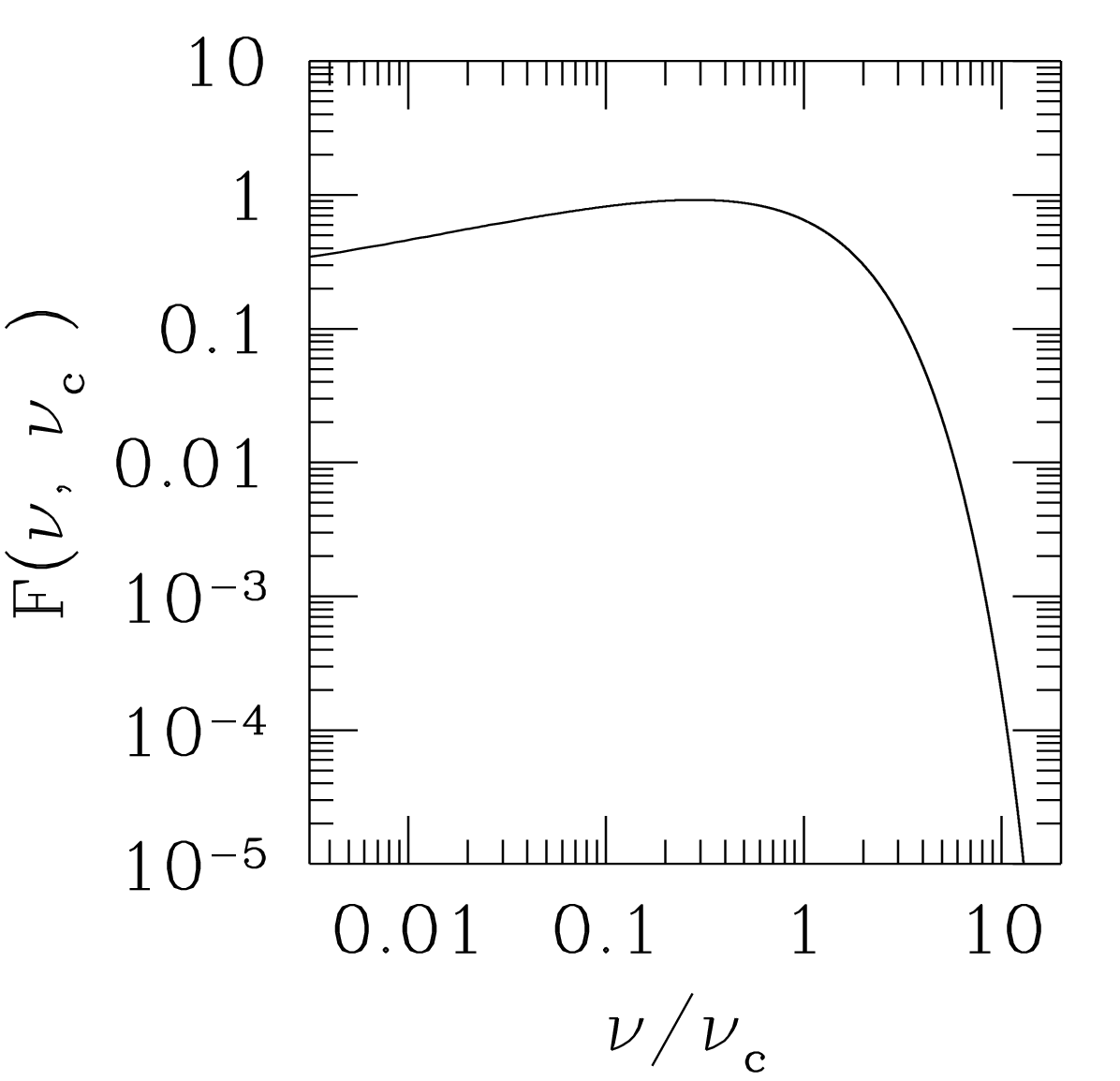}
\caption{The spectral distribution function of synchrotron radiation
from a monoenergetic electron
of Lorentz factor $\gamma$, where $\nu_{\rm c}$ is defined in 
(\ref{eq:synfreq}).  The spectrum is plotted linearly on the left and
logarithmically on the right}
\label{fig:figbess}      
\end{figure}

\noindent
where $K_{5/3}$ is the modified Bessel function of order 5/3.  The
spectrum peaks at $0.29\nu_{\rm c}$, as shown in
Fig.~\ref{fig:figbess}.  The synchrotron spectrum from a distribution
of electrons with some maximum Lorentz factor, $\gamma_{\rm max}$,
will therefore fall exponentially at high frequencies.  The gradual
turn-down at low frequencies, $\propto \nu^{1/3}$, is not expected to
be seen in practice.  Instead, for a homogeneous emitting region the
low-energy fall off will have a slope $\propto \nu^{2.5}$ due to the
source having become optically thick and suffering from synchrotron
self-absorption.

The luminosity per unit frequency for a number spectrum of electrons
$N(\gamma) d\gamma$ is

\begin{equation}
L_\nu = 2 \pi \sqrt{3}\,c\,m_{\rm e}\,r_{\rm e}\,\nu_{\rm g} \sin\alpha
\int F(\nu, \nu_{\rm c})\, N(\gamma) d\gamma
\label{eq:synlum}
\end{equation}

\noindent
where $r_{\rm e}$ is the classical electron radius, and this form is
valid in all systems of units.  For a power-law number distribution
of electrons $N(\gamma)\,d\gamma = \kappa\gamma^{-p}\,d\gamma$, an
analytical result can be found for frequencies satisfying
$\gamma_{\rm min}^2\nu_{\rm g} \ll \nu \ll \gamma_{\rm max}^2\nu_{\rm
g}$

\begin{equation}
L_\nu = \kappa \nu^{-(p-1)/2}\nu_{\rm g}^{(p+1)/2} \sin\alpha^{(p+1)/2}
m_{\rm e}\,c\,r_{\rm e} {3^{p/2} 2\pi\over p+1} \Gamma\left({p\over 4} + {19 \over
12}\right) \Gamma\left({p\over 4} - {1 \over
12}\right)
\label{eq:synlumpowerlaw}
\end{equation}

\noindent
where $\Gamma$ is the gamma function, and, again this form
is valid in all systems of units.  If the pitch angle distribution
is isotropic, the term in $\sin\alpha$ may be replaced by
$\sqrt{\pi}\Gamma[(p+5)/4]/2\Gamma[(p+7)/4]$.  The result for more
complicated electron spectra must be computed numerically using
(\ref{eq:synlum}).  

It is usual to use the symbol $\alpha$ for the
negative exponent of the power-law radiation spectrum ($L_\nu \propto
\nu^{-\alpha}$), where $\alpha = (p-1)/2$.  Particle acceleration by
ultra-relativistic shocks, likely to be important in jets, produces $p
\approx 2.2-2.3$ \cite{achterberg}, and $p$ will be steeper in
regions where energy losses are important.

\subsection{Inverse Compton scattering}
\label{sec:compton}

\begin{figure}[t]
\centering
\includegraphics[height=5cm]{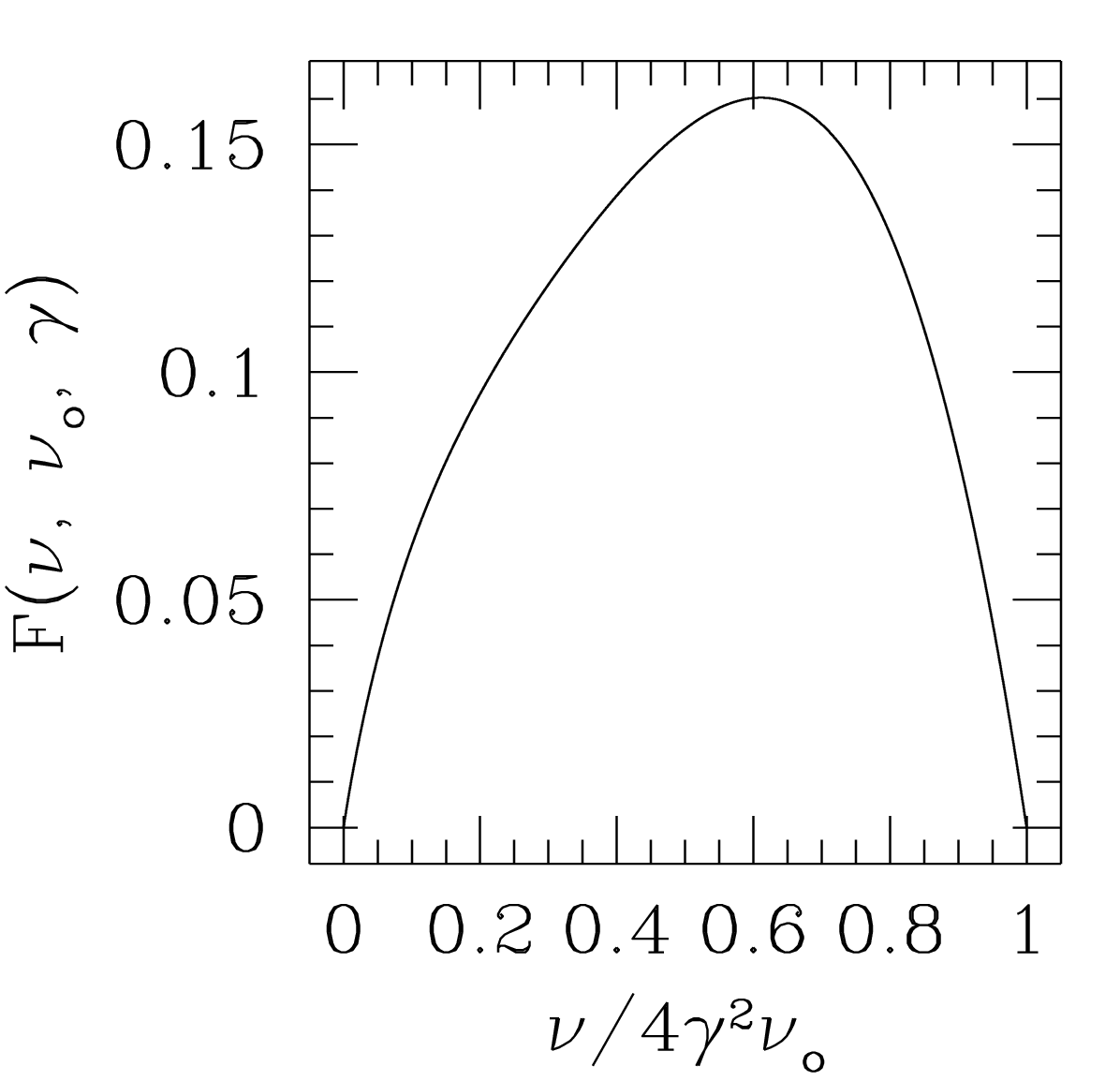}
\includegraphics[height=5cm]{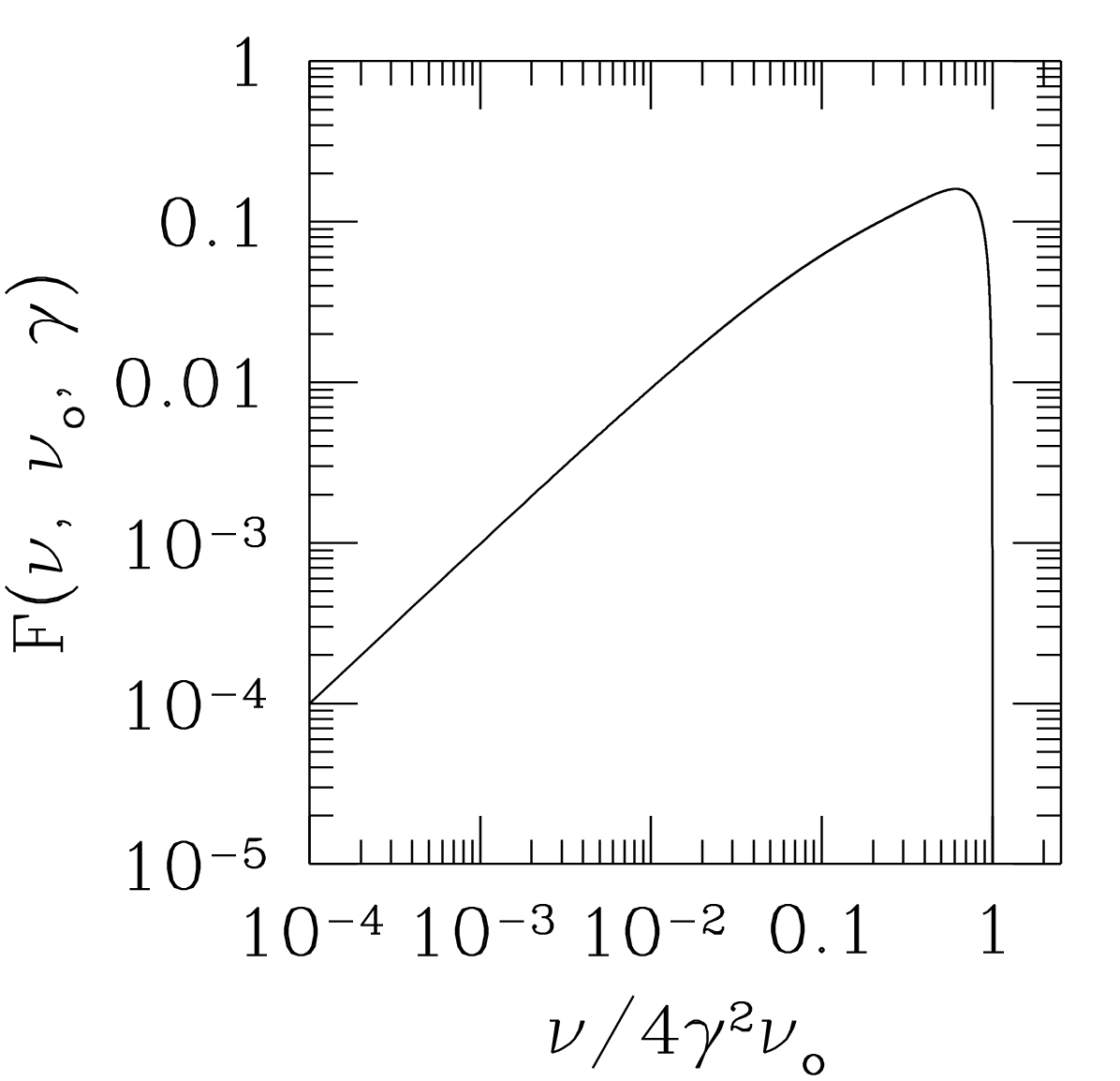}
\caption{The spectral distribution function of inverse Compton radiation
from a monoenergetic electron
of Lorentz factor $\gamma$ in an isotropic radiation field of
frequency $\nu_{\rm o}$ in the Thomson limit (\ref{eq:thompson}).
 The spectrum is plotted linearly on the left and
logarithmically on the right}
\label{fig:figfx}     
\end{figure}

Relativistic electrons lose energy by scattering photons to higher
energy in a process analogous to synchrotron radiation, where virtual
photons are replaced with real ones.  Providing an electron of mass
$m_{\rm e}$ and Lorentz factor $\gamma$ is scattering a photon of
low enough energy such that 

\begin{equation}
\gamma h\nu_{\rm o} \ll m_{\rm e}c^2
\label{eq:thompson}
\end{equation}

\noindent
the Thomson cross section, $\sigma_T$ is applicable, and the rate of loss
of energy of the electron in an isotropic radiation field of total
energy density $u_{\rm rad}$ is 

\begin{equation}
-dE/dt = (4/3) \sigma_{\rm T} c \gamma^2 \beta^2 u_{\rm rad} 
\label{eq:icloss}
\end{equation}

\noindent
where other symbols are as defined in Sect.~\ref{sec:synchrotron}.
The radiative lifetime can be calculated using (\ref{eq:tausyn}).
Where electrons are losing energy both by synchrotron and inverse
Compton emission, the ratio of total luminosity in the two emissions
is simply $L_{\rm iC}/L_{\rm syn} =  u_{\rm rad}/ u_{\rm B}$.

An exact result \cite{blum-gould} exists for the spectral
distribution function for a
monoenergetic electron of Lorentz factor $\gamma$ scattering an
isotropic radiation field of photons of frequency $\nu_{\rm o}$ up to
frequency $\nu$.  It is convenient to
define $X = \nu/4\gamma^2 \nu_{\rm o}$, where $(1/4\gamma^2) \leq X
\leq 1$, and then the spectral distribution function is proportional
to

\begin{equation}
F(\nu, \nu_{\rm o}, \gamma) = X f(X) = X (1 + X - 2X^2 + 2X \ln X)
\label{eq:fx}
\end{equation}

\noindent
This is plotted in Fig.~\ref{fig:figfx}.  The luminosity per unit
frequency for a number spectrum of electrons $N(\gamma)d\gamma$ and a
spectral number per unit volume of photons $n(\nu_{\rm o})d\nu_{\rm
o}$  is

\begin{equation}
L_\nu = 3 h\,c\,\sigma_{\rm T} \int \int 
n(\nu_{\rm o})\,N(\gamma)\,F(\nu, \nu_{\rm o}, \gamma)\,d\nu_{\rm o}\,d\gamma
\label{eq:iclum}
\end{equation}

\noindent
The relationship between the electron power-law index,
$p$, and the spectral index of inverse Compton radiation, $\alpha$,
is the same as for synchrotron radiation.
Since the mean value of X,

\begin{equation}
\int X f(X) dX /\int f(X) dX = 1/3
\label{eq:xmean}
\end{equation}

\noindent
the mean frequency of photons scattered in an isotropic radiation
field is given by

\begin{equation}
\nu = (4/3) \gamma^2 \nu_{\rm o}
\label{eq:icnumean}
\end{equation}

\begin{figure}[t]
\centering
\includegraphics[height=5.3cm]{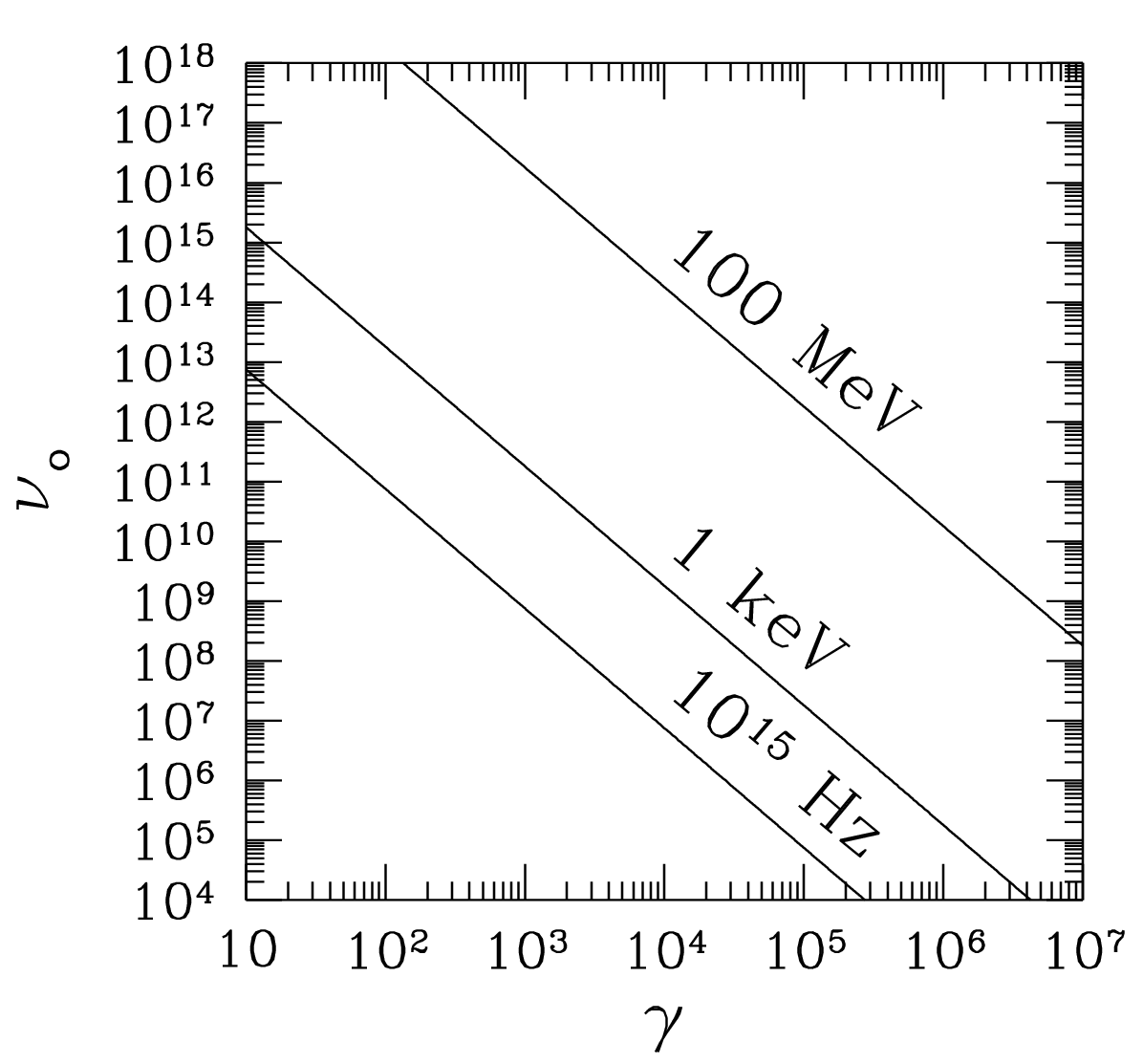}
\includegraphics[height=5.3cm]{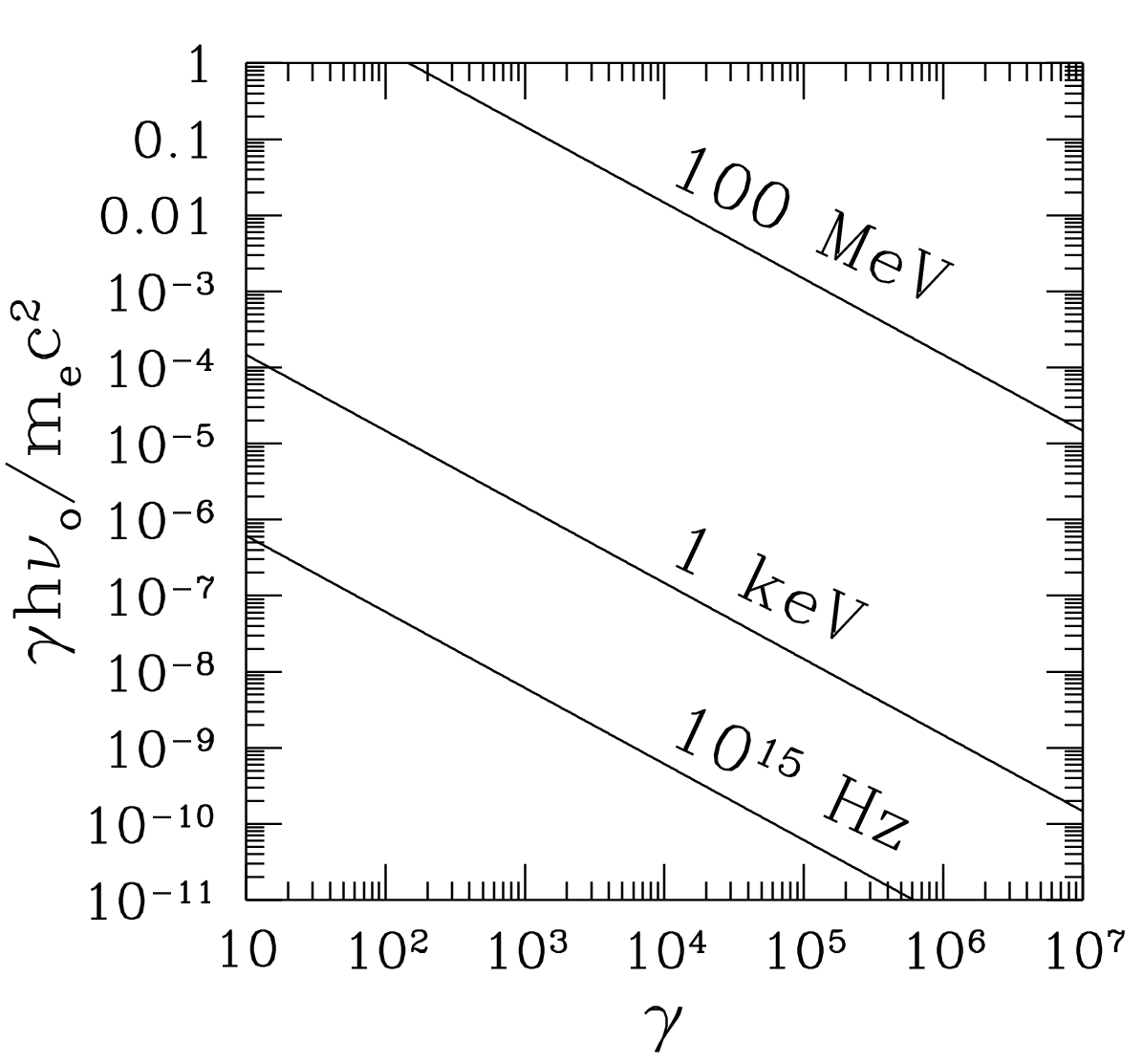}
\caption{The combinations of electron Lorentz factors and
photon frequencies which will produce photons of typical
optical, X-ray and $\gamma$-ray energies via inverse Compton
scattering of an isotropic photon distribution in
the Thomson limit is shown on the left.
The Thomson limit only holds if $\gamma h \nu_o/m_{\rm e} c^2 \ll 1$.
The right-hand plot shows this is satisfied for optical and X-ray
photons, but for $\gamma$-ray and higher energies the
Klein-Nishina cross-section begins to become important
}
\label{fig:syn}  
\end{figure}

Fig.~\ref{fig:syn} plots combinations of $\gamma$ and
$\nu_{\rm o}$ that will produce optical, X-ray, and $\gamma$-ray
photons according to (\ref{eq:icnumean}), and shows that in
most situations (\ref{eq:thompson}) is applicable.  As
(\ref{eq:thompson}) becomes violated, the Klein-Nishina cross section
must be used in place of $\sigma_{\rm T}$, as in these situations the
electron begins to lose a significant fraction of its energy in a
single scattering, e.g., \cite{blum-gould}.

The photon fields available to the electrons in jets are the
synchrotron radiation they have produced, in which case the emission
is known as synchrotron self-Compton (SSC) radiation, 
the cosmic microwave background (CMB)
which scales with redshift as $(1 + z)^4$, and photons from the
active-galaxy nucleus.  In practice the photon fields are rarely
expected to be isotropic when the effects of the geometry and bulk
speed of the jet plasma are taken into account.

\subsection{Thermal radiation and energy loss}
\label{sec:thermal}

For hot X-ray emitting gas containing ions of charge $Z_{\rm i}$, and
where the cooling in line radiation is unimportant, the X-ray
emissivity, $\cal E_\nu$ depends on temperature, $T$, and electron and
ion number densities, $n_{\rm e}$, $n_{\rm i}$, as

\begin{equation}
  {\cal E}_\nu = {32 \pi \over 3} \, 
                \left( {2 \pi \over 3} \right)^{1/2} \, 
                Z_{\rm i}^2 \, g_{\rm ff} \,
                n_{\rm i}\, n_{\rm e}\,
                m_{\rm e}\, c^2\, r_{\rm e}^3 \,
                \left( {m_{\rm e} c^2 \over k T} \right)^{1/2} \,
                e^{- h \nu / k T}
\label{eq:bremssapprox}
\end{equation}

\noindent
for all systems of units, where $r_{\rm e}$ is the classical electron radius.
$g_{\rm ff}(\nu, T, Z)$, the free-free Gaunt factor which accounts for
quantum-mechanical effects, is of order unity and a weak function of
frequency.    An approximate form \cite{rlightman} in the
case of most X-ray interest,  
$(kT/{\rm eV}) > 13.6 Z_{\rm i}^2$ and $Z_{\rm i} \leq  2$, is

\begin{equation}
  g_{\rm ff} = \left\{
    \begin{array}{ll}
       \left( {3 \over \pi} {k T \over h\nu} \right)^{1/2} 
        & \qquad \mbox{$h\nu > k T$} \\
       {\sqrt{3} \over \pi} \ln \left( {4 \over \zeta} {k T \over
    h\nu}
         \right) 
        & \qquad \mbox{$h\nu < k T$} 
    \end{array}
  \right.
\end{equation}

\noindent
where the constant  $\zeta = 1.781$. For heavier ions more
complicated forms must be used, and since the heavier ions contribute
disproportionately to the X-ray output (because of the $Z_{\rm i}^2$
factor), the calculation of the correct Gaunt factor becomes a
computational issue.

In SI units, (\ref{eq:bremssapprox}) becomes

\begin{equation}
{\cal E}_\nu = 6.8 \times 10^{-51} Z_{\rm i}^2 T^{-1/2} 
(n_{\rm e}/{\rm m^{-3}})\,(n_{\rm i}/{\rm m^{-3}})\,
g_{\rm ff}\,e^{-h\nu/kT} \quad{\rm W~m}^{-3}~{\rm Hz}^{-1}
\label{eq:bremssapproxsi}
\end{equation}

\begin{figure}[t]
\centering
\includegraphics[height=7cm]{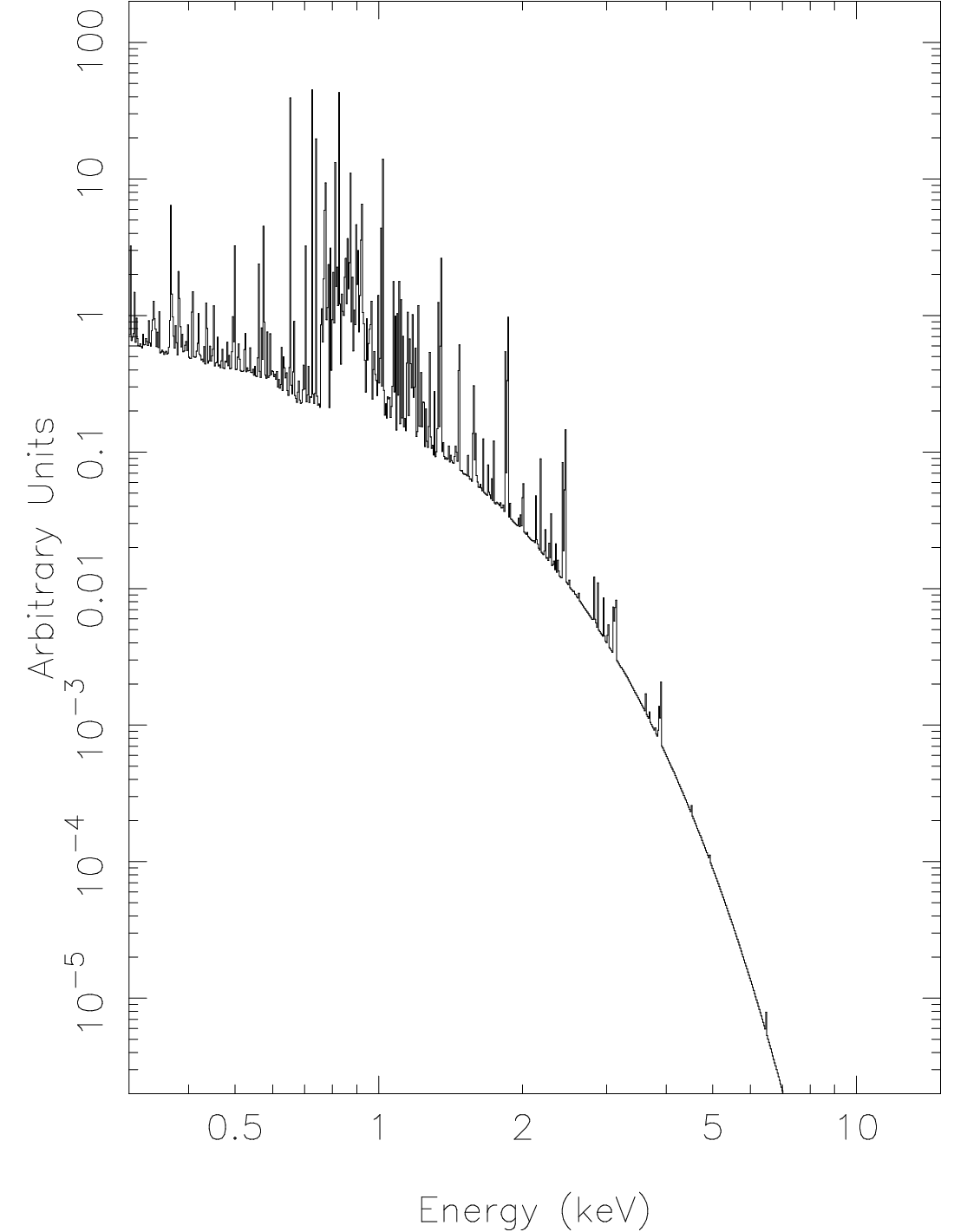}
\includegraphics[height=7cm]{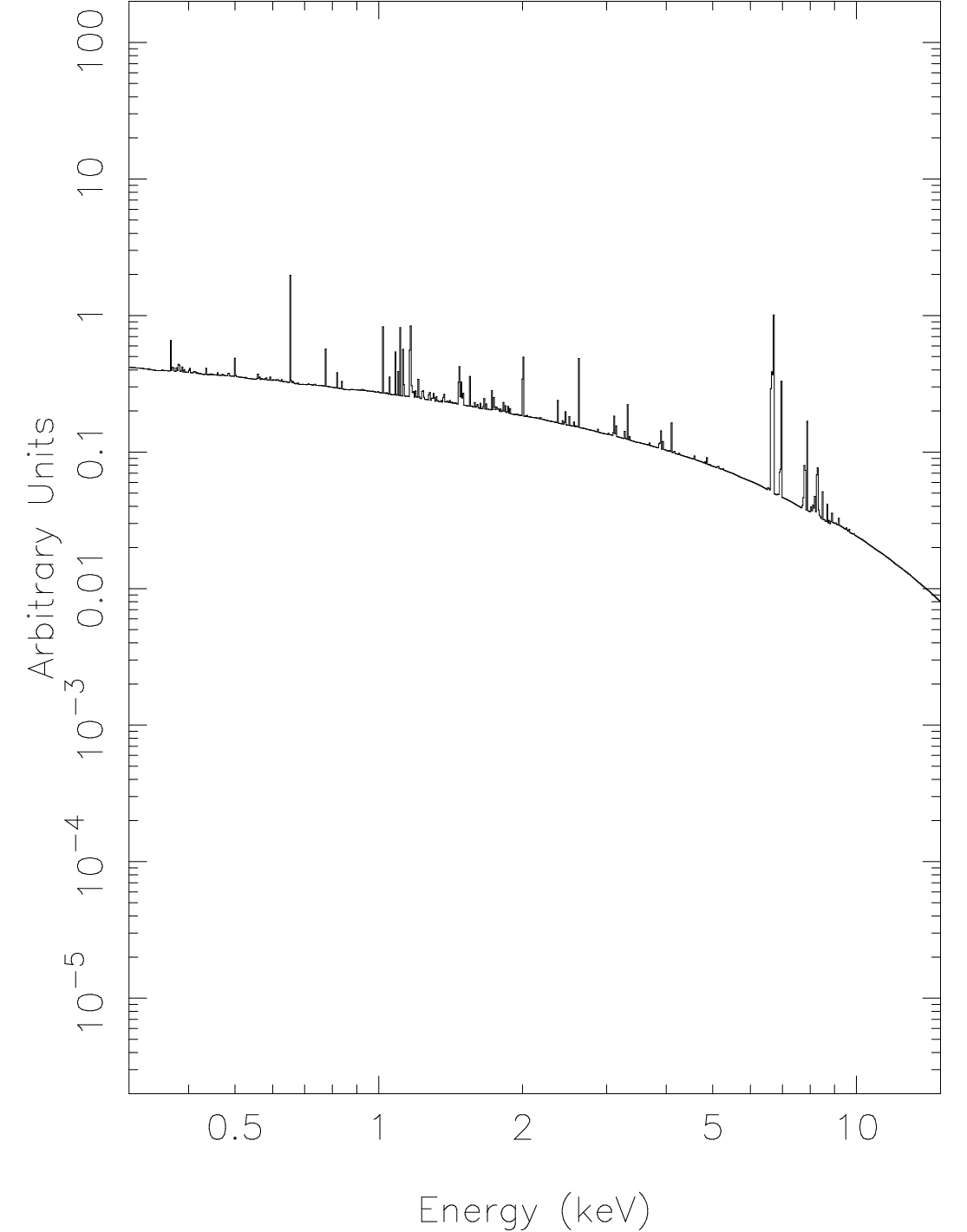}
\caption{Emissivity distributions of gas of $kT = 0.5$ keV (left) and
$kT = 5$ keV (right) with normal cosmic abundances from the
\texttt{APEC} model in \texttt{XSPEC}.  The line emission,
superimposed on the thermal bremsstrahlung continuum, is a more
prominent channel for cooling in cool gas.  The resolution of these
model spectra is higher than that of current X-ray detectors }
\label{fig:figapec}    
\end{figure}

At temperatures below $\sim 2$~keV, line radiation cannot be ignored,
and the plasma models incorporated into X-ray spectral-fitting
programs such as \texttt{XSPEC} \cite{arnaud-xspec} can be used to
find ${\cal E}_\nu$. Examples for plasma of $kT = 0.5$~keV and 5.0~keV
are shown in Fig.~\ref{fig:figapec}.  It is common to define the rate
of loss of energy per unit volume per unit frequency per $n_{\rm
p}n_{\rm e}$ as $\Lambda(\nu)$, where $n_{\rm p}$ is the proton number
density.  In \texttt{XSPEC}, $\Lambda(\nu)$ is normalized by
$(1+z)^2\int n_{\rm p}n_{\rm e} dV/4\pi D_{\rm L}^2$ in fitting to
data, where $V$ is volume and $D_{\rm L}$ is luminosity distance.
This allows $n_{\rm p}$ to be determined assuming some geometry for
the source, as illustrated in Sect.~\ref{sec:medium}.

The total energy-loss rate per unit mass is given by

\begin{equation}
\epsilon =  {X\int {\cal E}_\nu d\nu \over m_{\rm H} n_{\rm p}}
\label{eq:thermalratepermass}
\end{equation}

\noindent
where $X$ is the mass fraction in hydrogen, which is 0.74 for normal
cosmic (i.e., solar) abundances.

\begin{figure}[t]
\centering
\includegraphics[height=8cm]{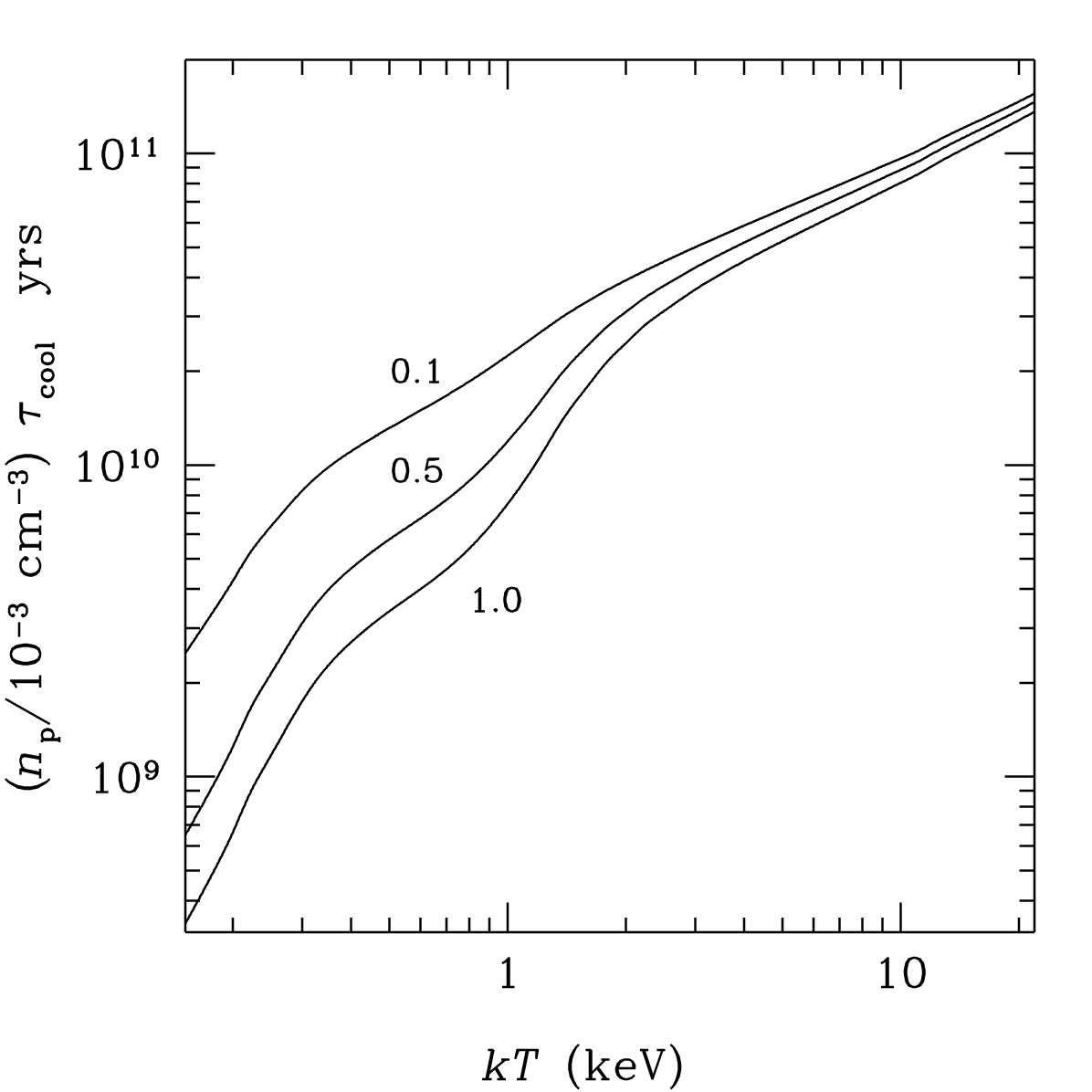}
\caption{Curves of $\tau_{\rm cool}n_{\rm p}$ (with $n_{\rm p}$ in units of
$10^3$ m$^{-3}$ or $10^{-3}$ cm$^{-3}$) as a function of $kT$
for three metal abundances with respect to solar: 0.1, 0.5, 1.0
}
\label{fig:cooling} 
\end{figure}

If the energy loss is
sufficiently rapid, the radiating gas will fall towards the centre of
the gravitational potential well as its pressure support fails. The
rate of infall is therefore determined by whether the characteristic
timescale for radiative energy loss is greater than or less than the
timescale for energy input, perhaps from the dissipation of energy by
the radio jet.

If the timescale for energy loss is defined by the logarithmic rate of
change of temperature in the comoving fluid frame,

\begin{equation}
  \tau_{\rm cool} = - \left( {D \ln T \over Dt} \right)^{-1}
\end{equation}

\noindent
where $D/Dt$ is the convective derivative (${\partial / \partial t} +
\vec{v}\cdot\vec{\nabla}$), which describes the rate of change in a
frame moving at velocity $\vec{v}$ with the gas, then the value of
$\tau_{\rm cool}$ can be obtained from the entropy equation

\begin{equation}
  {D s \over Dt} = - {\epsilon \over T}
\end{equation}

\noindent
where $s(T, P)$ is the
entropy per unit mass of the gas and $P$ is the pressure.
For isobaric ($P$ = constant) losses in a monatomic gas,

\begin{equation}
  \tau_{\rm cool} = {5 \over 2} {k T \over \epsilon\,\mu 
    m_{\rm H}}
\end{equation}

Curves of $\tau_{\rm cool}n_{\rm p}$ as a function of $kT$,
calculated using the \texttt{APEC} plasma model
in \texttt{XSPEC}, are shown
in Fig.~\ref{fig:cooling}.  At high temperatures, $\tau_{\rm cool}
\propto T^{1/2}$, but at $T \leq 2$~keV line emission from
common metal species in the gas becomes important, and so $\epsilon$
and $\tau_{\rm cool}$ become strong functions of $kT$.  The offset
between the curves at large $kT$ arises from the increased density of
highly-charged ions in the radiating plasma as the metal abundance
increases.

\section{Effects of bulk relativistic motion}
\label{sec:doppler}

The first evidence for bulk relativistic motion in jets was on pc scales.
In the early 1970s, the technique of Very Long Baseline
Interferometry (VLBI) became established, and the first detections
of apparent superluminal motion of small-scale radio components were
being made (see \cite{pearson-zensus} for a short review).
After some years of debate, it became widely accepted that the true
explanation of the effect was as proposed earlier, \cite{rees}, and
that the apparently superluminal velocities
resulted from components moving towards
the observer at bulk relativistic speeds.

From a consideration of light travel time, it is straightforward to
show that a source moving with a relativistic speed $\beta c$, bulk Lorentz
factor $\Gamma (=1/\sqrt{1-\beta^2})$, at angle $\theta$ towards the observer, 
has an apparent transverse velocity of 

\begin{equation}
v_{\rm app} = \beta c
\sin\theta/(1 - \beta\cos\theta)
\label{eq:vapp}
\end{equation}

\noindent
which takes on a maximum
value of $\Gamma\beta c$ when $\sin\theta = 1/\Gamma$.
Hence large $\beta$ ($>1/\sqrt{2}$) and small $\theta$ can easily result in $v_{\rm app}
> c$.

The  bulk relativistic Doppler factor is defined as

\begin{equation}
\delta = {1/ \Gamma ( 1 - \beta \cos\theta)}
\label{eq:doppler}
\end{equation}

\noindent
and takes on a large value for large $\beta$, small $\theta$.  For a
spherical blob of emission, the apparent spectral luminosity, $L_\nu$,
is increased by $\delta^{3 + \alpha}$, where $\alpha$ is the spectral
index (Sect.~\ref{sec:synchrotron}), although in most steady jets
$\delta^{2 + \alpha}$ is more appropriate \cite{phinney,
scheuer-readhead}.

A variable source will show intensity changes on timescales shorter by
$\delta$ than the true value, and beaming effects can have a dramatic
effect on model parameters.  For example, it was shown that the
predicted X-ray flux density from SSC in a compact, spherical,
variable radio source must be reduced by a factor of
$\delta^{2(3\alpha+5)}$ if variability is used to measure the size,
and the observed self-absorption turn-over frequency is used to
estimate the magnetic-field strength \cite{marscher,
marscher-bursting}.

Relativistic beaming has a major effect on a jet's appearance, and by
the 1980s the first models to unify classes of active galaxies through
jet orientation were proposed, e.g., \cite{orr-browne,
scheuer-readhead}, and subsequently further developed to account for
multiwavelength properties, e.g., \cite{barthel, scheuer,
urry-padovani}.  It is common to assume that a radio source has two
oppositely-directed jets that are intrinsically the same, although
there is doubt that this need be the case either from jet-production
considerations or asymmetries in the environment through which the
jets propagate.  Under the assumption of intrinsic similarity, the
brightness ratio between the approaching and receding jets is given by

\begin{equation}
R_{\rm J} = \left({1 - \beta\cos\theta \over 1 +
\beta\cos\theta}\right)^{-(\alpha + 2)}
\label{eq:jetcounterjet}
\end{equation}

This constraint on $\beta$ and $\theta$ is often used in conjunction
with a constraint on $v_{\rm app}$
to estimate separately the jet speed and the angle to the line of
sight, e.g., \cite{jones-6251, worr-3c48}.  The core dominance, $R_{\rm cd}$, defined as the
flux density of beamed emission divided by that of unbeamed extended
emission, is also commonly used as an indicator of the
angle of a source to the line of sight.  When applied to a twin-jet
source, e.g., \cite{orr-browne},

\begin{equation}
R_{\rm cd} = {R_{{\rm cd}_{(\cos\theta=0)}} \over 2}
\left[(1 - \beta\cos\theta)^{-(\alpha + 2)} +
(1 + \beta\cos\theta)^{-(\alpha - 2)}\right]
\label{eq:twinjet}
\end{equation}

Bulk relativistic motion towards the observer has an important effect
on inverse-Compton scattering in the case of an external photon field
(i.e., not SSC) \cite{dermer}.  In its rest frame an electron now sees
a directionally enhanced radiation field and the rate of scatterings
increases. If the photon field is isotropic in the observer's frame
(such as the CMB), the increase in scattering rate is by a factor of
$\delta$.  Since lower-energy electrons (of which there are more) are
needed to scatter photons of a given rest-frame frequency to the
observed frequency, there is a further enhancement by $\delta^\alpha$.
This means that the observed luminosity at a given frequency from
inverse-Compton scattering of the CMB will be increased relative to
the synchrotron luminosity by $\delta^{1+\alpha}$, assuming that the
magnetic field and electron spectrum are held constant.  Although this
in itself is not a large factor compared with $\sim \delta^{2+\alpha}$
(see above) that arises from any radiation mechanism and applies both
to the synchrotron and inverse-Compton emission, it can be sufficient
to raise the ratio of X-ray to radio emission enough to fit
observations of quasar jets (Sect.~\ref{sec:speed}).  For a source
radiating at minimum energy (Sect.~\ref{sec:bmin}), while the X-ray
emission from inverse-Compton scattering of CMB photons is increased,
the X-ray emission from SSC is reduced.  To increase this ratio via
the SSC mechanism requires taking the source out of minimum energy
through adding relativistic electrons.  A beamed source has
intrinsically lower total energy content, and, despite the higher jet
speed increasing the kinetic energy, it may be in a condition of
minimum total (radiative and kinetic) power \cite{ghisellini-celotti}.

\section{The external medium}
\label{sec:medium}

Often the bulk of the X-ray emission from an active galaxy is not jet
related, but rather arises from the hot interstellar, intergalactic,
or intracluster medium around the jet. If we assume that this gas is
almost at rest in the local gravitational potential well, with little
transfer of energy or matter to or from it, then it will be in a state
close to hydrostatic equilibrium.

Under these circumstances, the gas will take up a density and
temperature profile which depend on the distribution of mass and the
thermal history of the gas. The principal governing equation is that
of hydrostatic equilibrium, with 

\begin{equation}
 \vec{\nabla} P = -\rho \, \vec{\nabla} \Phi
\label{eq:med1}
\end{equation}

\noindent
where $\Phi$ is the gravitational potential and $\rho$ is the total
mass density of the gas. We may assume that the gas has the ideal gas
equation of state

\begin{equation}
 P = {\rho \, k \, T \over \mu \, m_{\rm H}}
= {n_{\rm p}\, k\,T \over X\,\mu}
\label{eq:med2}
\end{equation}

\noindent
to relate $P$ and $\rho$, where $\mu$ is the mass per particle in
units of $m_{\rm H}$, given by

\begin{equation}
 \mu = \left( 2 X + {3 \over 4}Y + {1 \over 2} Z \right)^{-1} =
       0.60
\label{eq:med3}
\end{equation}

\noindent
for a gas with solar abundances ($X = 0.74$, $Y = 0.25$, $Z
= 0.01$). 

If the mass distribution is spherically symmetrical, then 

\begin{equation}
 {d \Phi \over dr} = {G \, M_{\rm tot}(r) \over r^2}
\label{eq:med4}
\end{equation}

\noindent
where $M_{\rm tot}(r)$ is the total mass within radius $r$, and so
the mass density and temperature profiles of the gas are related to the
distribution of total mass by

\begin{equation}
  G \, M_{\rm tot}(r) = - {k T r \over \mu m_{\rm p}}
  \, \left( {d\ln\rho \over d\ln r} + {d\ln T \over d\ln r}
  \right) \quad .
  \label{eq:equilmass}
\end{equation}

A consistent solution of this equation is obtained by assuming that
the gas is isothermal and that the mass distribution follows 

\begin{equation}
  M_{\rm tot}(r) = 2 \, M_{\rm c} {r^3 \over r_{\rm c} \left(
                        r^2 + r_{\rm c}^2 \right)}
 \label{eq:mtot}
\end{equation}

\noindent
where $r_{\rm c}$, the core radius, defines the characteristic scale
of the mass distribution and $M_{\rm c}$ is
the mass within $r_{\rm c}$.
The density of the gas then follows
the so-called isothermal $\beta$ model \cite{cavaliere} 

\begin{equation}
 \rho = \rho_0 \left( 1 + {r^2 \over r_{\rm c}^2}
             \right)^{-{3 \over 2}\beta} 
 \label{eq:beta}
\end{equation}

\noindent
where $\beta$ is a constant which determines the shape of the gas
distribution, and depends on the ratio of a
characteristic gravitational potential to the thermal energy
per unit mass in the gas 

\begin{equation}
  \beta = {2 \over 3} \, {\mu m_{\rm H} \over k T} 
                      \, {G M_{\rm c} \over r_{\rm c}} \quad .
\label{eq:med5}
\end{equation}

The alternative derivation of (\ref{eq:beta}) given by
\cite{cavaliere} brings out the interpretation of $\beta$ in
terms of the relative scale heights or, equivalently, the ratio of
energy per unit mass in gas and dark matter.
$\rho_0$, $r_{\rm c}$, and $\beta$ in (\ref{eq:beta}) are often
fitted to provide convenient measures of the gas distribution's mass,
scale and shape without considering the detailed properties of the
underlying mass distribution (\ref{eq:mtot}).

It should be noted that the physical consistency of this much-used
model for the gas distribution depends on radial symmetry (a simple 
distortion into an ellipsoidal model for the gas density $\rho$ would
imply a mass distribution which is not necessarily positive
everywhere), and on gas at different heights in the atmosphere having
come to the same temperature without having necessarily followed the
same thermal history.  This is problematic given the various origins
of gas --- some will have fallen into the
cluster, some will have been ejected from stars, and some may have
been moved by the jet. The solution (\ref{eq:beta}) is not unique in
the sense that gas with a different thermal history might follow a
significantly different density distribution. Thus, for example, if
the gas has the same specific entropy at all heights, and so is
marginally convectively unstable, with

\begin{equation}
 P \propto \rho^{\gamma}
\end{equation}

\noindent
with $\gamma$ being the usual specific heat ratio ($5 /3$ for a
monatomic gas), the mass distribution (\ref{eq:mtot}) would lead
to a gas density distribution 

\begin{equation}
  \rho^{\gamma-1} = \rho_0^{\gamma-1} \, \left( 1 - \beta_{\rm A} \ln \left
                    ( 1 + {r^2 \over r_{\rm c}^2} \right) \right)
\end{equation}

\noindent
where $\beta_{\rm A}$ is a structure constant with a similar meaning to
$\beta$, and the relation only applies within the radius at which
$\rho \rightarrow 0$.

A similar procedure for an NFW mass profile \cite{navarro} and an
isothermal gas leads to a density distribution

\begin{equation}
 \rho = \rho_{\rm s} \, 2^{-\alpha} \, \left( 1 + {r \over r_{\rm
         s}} \right)^{ \alpha r_{\rm s} / r}
  \label{eq:nfw}
\end{equation}

\noindent
where the new structure constant is

\begin{equation}
 \alpha = {1 \over \ln 2 - {1 \over 2}} \, {\mu m_{\rm H} \over k T}
   \, {G M_s \over r_s} 
\end{equation}
($r_s$ is the scale of the NFW model and $M_s$ is the mass
within radius $r_s$).  It can be seen that in this solution $\rho \rightarrow
\infty$ as $r \rightarrow 0$, since the NFW density profile has a cusp
at $r = 0$.

The total masses for either the NFW profile or (\ref{eq:mtot}) diverge
as $r \rightarrow \infty$, so both profiles must be truncated at some
radius, such as $r_{\rm 200}$, the radius at which the mean enclosed
mass density is $200 \times$ the critical density of the Universe,
$\rho_{\rm crit}$ at redshift $z$. Care should be taken in using the
solutions for $\rho$ to ensure that the fraction of mass in gas does
not exceed the cosmological bound ($\sim 17\%$).

The run of density and temperature in the atmosphere around a jet are
measured from the X-ray image and spectrum, where a density model for
the gas of the form of (\ref{eq:beta}) or (\ref{eq:nfw}) is fitted to
the X-ray surface brightness. The X-ray surface brightness at
frequency $\nu$ at a point offset by $r$ in {\it projected} distance
from the centre of a spherical gas distribution is (in
energy per unit time per unit solid angle per unit frequency)

\begin{equation}
  \Sigma_\nu(r) d\Omega = (1 + z) \, {\int \, n_{\rm e} n_{\rm p} 
     \Lambda_\nu(T) dV \over 4 \pi D_{\rm L}^2}
\end{equation}

\noindent
where $\Lambda_\nu$ is the emissivity of the gas at frequency $\nu$,
and $D_{\rm L}$ is the luminosity distance of the gas. The $(1+z)$
factor takes account of the redshifting of time and frequency
in the definition of
$\Sigma_\nu$. The volume element in the integral is 

\begin{equation}
  dV = dl D_{\rm A}^2 d\Omega
\end{equation}

\noindent
where $dl$ is an element of distance along the line of sight, $D_{\rm
A}$ is the angular diameter distance, and $d\Omega$ is the element of
solid angle, so that the surface brightness becomes

\begin{equation}
  \Sigma_\nu(r) = {\int \, n_{\rm e} n_{\rm p} \Lambda_\nu(T) dl
                      \over 4 \pi (1 + z)^3 } \quad .
\end{equation}

\noindent
For an isothermal gas with a $\beta$-model density distribution

\begin{equation}
  \Sigma_\nu(r) = {\Lambda_\nu(T) \, n_{\rm e0} \, n_{\rm p0}
                      \over 4 \, \pi \, (1 + z)^3 } \,
                      \int \left( 1 + {r^2 + l^2 \over r_{\rm c}^2}
                      \right)^{-3 \beta} \, dl
\end{equation}

\noindent
where the integral is taken along the line of sight with limits of
$\pm \infty$ or to some cut-off radius. In the former
case the integral is simple, and the X-ray surface brightness can be
written \cite{birk-n6251}

\begin{equation}
  \Sigma_\nu(r) = {\Lambda_\nu(T) \, n_{\rm e0} \, n_{\rm p0} \, r_c
                       \over 4 \, \pi \, (1 + z)^3 } \,
                       \left( 1 + {r^2 \over r_{\rm c}^2}
                       \right)^{{1 \over 2}-3 \beta} \,
                       \sqrt{\pi} \,
                       {\Gamma(3\beta - {1 \over 2}) \over
                       \Gamma(3\beta)}
\label{eq:betasb}
\end{equation}

\noindent
It is also useful to write down the observed flux density, $S_\nu$, of
a gas distribution out to angle $\theta$ from the centre, since this
is the quantity generally fitted in X-ray spectral analyses. This is

\begin{eqnarray}
  S_\nu(\theta) &=& \int_0^{\theta} \, 2 \pi \theta \, d\theta
                     \Sigma_\nu(r) \\
                &=& {\Lambda_\nu(T) \, n_{\rm e0} \, n_{\rm p0} \,
                     \theta_c^3 D_{\rm A} 
                     \over (1 + z)^3 } \,
                     {\sqrt{\pi} \over 4} \,
                     {\Gamma(3\beta - {3 \over 2}) \over
                     \Gamma(3\beta)} 
                     \left( 1 - \left( 1 + {\theta^2 \over \theta_{\rm
                     c}^2} \right)^{{3 \over 2}-3 \beta} \right)
\end{eqnarray}

\noindent
where $\theta_c$ is the angular equivalent of the core radius
$r_c$. Now, \texttt{XSPEC} (Sect.~\ref{sec:thermal}) calculates
a normalization, $\cal N$, for
any thermal gas model, where

\begin{equation}
  {\cal N} = {\int n_{\rm e} n_{\rm p} dV \over 4 \pi D_{\rm A}^2 (1 +
    z)^2} \equiv {(1+z) S_\nu \over \Lambda_\nu}
\end{equation}

\noindent
and this is related to the parameters of the isothermal $\beta$ model
by 

\begin{equation}
  {\cal N} = {n_{\rm e0} \, n_{\rm p0} \,
              \theta_c^3 D_{\rm L} 
              \over (1 + z)^4 } \,
              {\sqrt{\pi} \over 4} \,
              {\Gamma(3\beta - {3 \over 2}) \over
               \Gamma(3\beta)} 
               \left( 1 - \left( 1 + {\theta^2 \over \theta_{\rm
                 c}^2} \right)^{{3 \over 2}-3 \beta} \right) \quad 
\label{eq:betaspecnorm}
\end{equation}

The fitting process in \texttt{XSPEC} takes account of the form
of $\Lambda_\nu$ and the conversion from energy units ($S_\nu$)
to the count rate in energy bins used by X-ray detectors.

This expression shows how to use the normalization $\cal N$ found by
\texttt{XSPEC}, or some other code, from the X-ray spectrum 
within angular radius $\theta$ of the centre, and $\beta$ and $r_c$
as fitted from the radial profile of the X-ray emission, to measure
the central electron and proton densities which are related by

\begin{equation}
 {n_{\rm e} \over n_{\rm p}} =  {1 + X \over 2 X} \sim 1.18
\end{equation}

The proton density found in this way, and the gas temperature measured
from the spectrum, provide crucial information on the gas
environments of radio jets and hence the external forces on the jet.
The separation of X-rays from
the non-thermal processes associated with the radio source and the
thermal processes associated with the environment is therefore
an important part
of establishing the physics of a jet. This is particularly true for
low power (FRI) jets, which are in direct contact with the external
medium (Sec.~\ref{sec:fri}), but is also important for establishing
the balance of static and ram-pressures in powerful sources.

\section{Simple radio-jet models}
\label{sec:models}

The FRI and FRII classifications, into which extragalactic radio
sources were separated by \cite{fr}, remain in wide usage today.
Although based on radio morphology, \cite{fr} found a relatively sharp
division in radio luminosity, with most sources of total 178-MHz
luminosity below $2 \times 10^{25}$ W Hz$^{-1}$ sr$^{-1}$ being of FRI
classification, and the others of FRII.  Some dependence of the
location of the FRI/FRII boundary on the optical magnitude of the host
galaxy has also been found \cite{owen-ledlow}, suggesting the effects
of environment or the mass of the originating black hole are important
in determining the type of radio galaxy formed by an active nucleus.
However, the primary cause of the structural differences
is believed to be due to the speed of the primary jet fluid, with
the beams producing FRII sources being supersonic with respect to the
ambient medium, whereas FRI jets
are seen as the result of turbulent transonic or subsonic flows.
The following sub-sections describe the observational consequences.

\subsection{High-power FRII jets}
\label{sec:frii}

\begin{figure}[t]
\centering
\includegraphics[height=7.5cm]{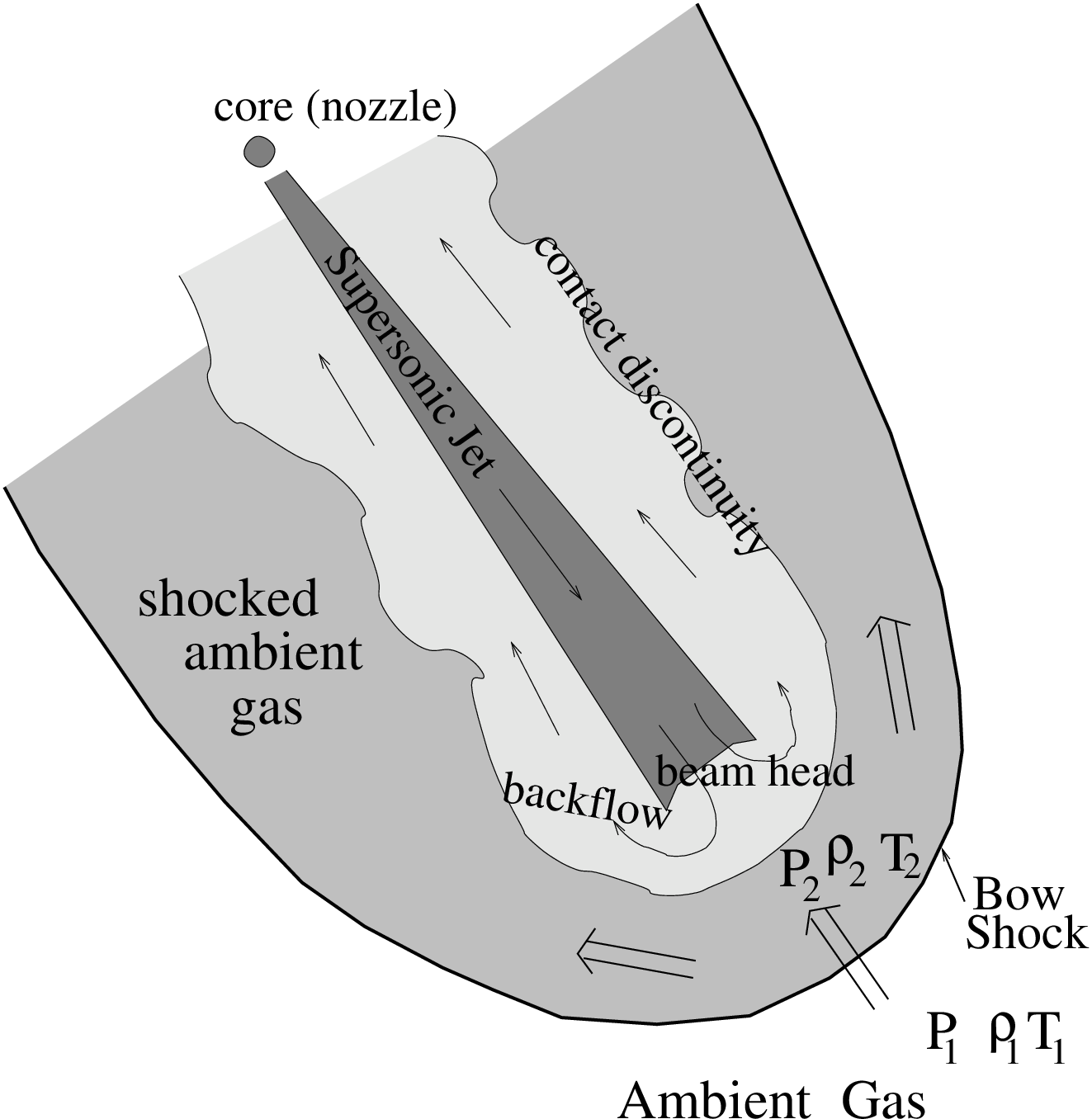}
\caption{In the standard model for powerful radio sources,
    a supersonic jet (dark grey) terminates at the beam head, producing a
    radio hotspot.
    Provided the shocked radio-emitting fluid forming the radio lobe
    (light grey) has enough internal energy or momentum density to drive
    a leading bow shock, ambient X-ray-emitting gas will be heated as
    it crosses the bow shock to fill the medium-grey region}
\label{fig:supersonic} 
\end{figure}

\begin{figure}[b]
\centering
\includegraphics[height=5.2cm]{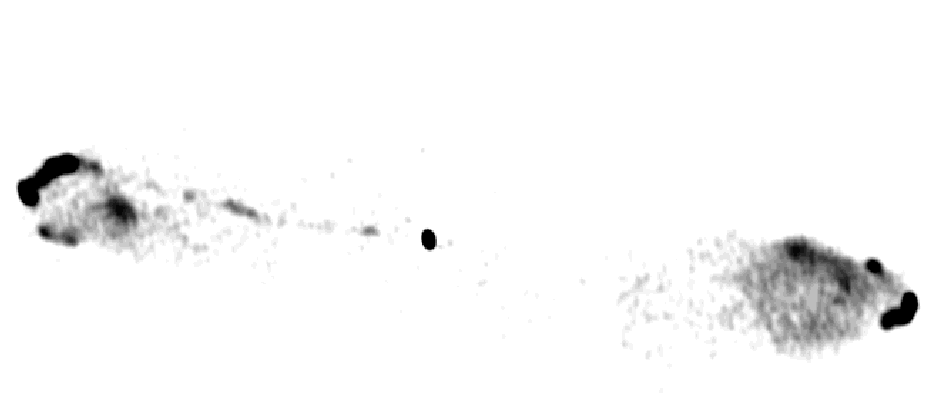}
\caption{8.4 GHz VLA A and B array radio image of an example
high-power FRII source,
the radio galaxy 3C 220.1.  The resolution is 0.3 $\times 0.2$
arcsec, and the image is from \cite{worr-c220.1}.
The angular length of the source is about 28 arcsec which, at
the galaxy's redshift of $z=0.61$, corresponds to 187 kpc
($H_o = 70$~km s$^{-1}$ Mpc$^{-1}$, $\Omega_{\rm m} = 0.3$,
$\Omega_\Lambda = 0.7$)
}
\label{fig:radio220}   
\end{figure}

The standard model for the expansion of a powerful radio source
powered by a jet which is supersonic with respect to the
X-ray-emitting interstellar medium (ISM) is illustrated in
Fig.~\ref{fig:supersonic}.  The jet terminates at the beam head (in a
feature identified as the radio hotspot) where the jet fluid passes
through a strong shock to inflate a cocoon of radio-emitting plasma.
The energy and momentum fluxes in the flow are normally expected to be
sufficient to drive a bow shock into the ambient medium ahead of the
jet termination shock.  In the rest frame of the bow shock, ambient
gas is heated as it crosses the shock to fill a region surrounding the
lobe of radio-emitting plasma.  
Observationally on kpc scales we see
the radio emission from well-collimated jets feeding edge-brightened
lobes.  An example is shown in Fig.~\ref{fig:radio220}.

The sound speed in gas of temperature $T$ is

\begin{equation}
c_{\rm s} = \sqrt{\gamma kT \over \mu m_{\rm H}}
\label{eq:csound}
\end{equation}

\noindent
where $\gamma$ is the ratio of specific heats ($\gamma$ = 5/3), $k$ is
the Boltzmann constant, $m_{\rm H}$ is the mass of the hydrogen atom,
and $\mu m_{\rm H} = 0.6 m_{\rm H}$ is the mass per particle.
Under these conditions,
$c_{\rm s} \approx 516 ({kT/{\rm keV}})^{1/2}$ km s$^{-1} \approx 0.54
({kT/{\rm keV}})^{1/2}$ kpc Myr$^{-1}$.

The Mach number of the speed of advance, $v_{\rm adv}$, of the bow
shock into the ambient medium is ${\cal M} = v_{\rm adv} / c_{\rm s}$,
which in convenient units can be expressed as

\begin{equation}
{\cal M} \approx  580 (v_{\rm adv}/c) (kT/{\rm keV})^{-1/2}
\label{eq:mach}
\end{equation}

\noindent
where $c$ is the speed of light.  
The advance of the bow shock is likely to be slow with respect
to the bulk speed of the jet.  However, 
the jets need not have bulk relativistic motion (Sect.~\ref{sec:doppler})
for the general description
of an FRII source to hold.

The state of the X-ray gas close to a radio galaxy with a leading bow
shock should reflect the source dynamics, and
in addition to the X-ray emission of the ambient medium
(Sect.~\ref{sec:medium}), we may also
expect to see hotter shocked gas surrounding the radio lobe
(Fig.~\ref{fig:supersonic}).
In a simple application of the
Rankine-Hugoniot conditions for a strong shock
\cite{spitzer}, the pressure, density, and temperature,
respectively, in the unshocked (subscript 1) and shocked (subscript 2)
regions at the head of the bow shock are related by

\begin{equation}
P_2/P_1 = (5 {\cal M}^2 -1)/4
\label{eq:rhpress}
\end{equation}

\begin{equation}
\rho_2/\rho_1 = 4 {\cal M}^2/ ({\cal M}^2 +3)
\label{eq:rhdens}
\end{equation}

\begin{equation}
T_2/T_1 = (5 {\cal M}^2 -1)({\cal M}^2 +3)/16 {\cal M}^2
\label{eq:rhtemp}
\end{equation}

\noindent
for a monatomic gas.

Using the above equations, and converting density into an
emissivity using \texttt{XSPEC}, as discussed in Sect.~\ref{sec:thermal},
we find that in the energy band 0.8--2 keV, where
{\it Chandra\/} and XMM-Newton are most sensitive, for
an ${\cal M} = 4$ shock, the X-ray emissivity contrast between shocked
and unshocked gas is a factor of 3 higher if the ambient gas is at
a galaxy temperature of $\sim 0.3$~keV than if the
external medium has a cluster temperature of $\sim 4$~keV. 

Complications apply in reality, and in practice these are difficult to
treat even with data from observatories as powerful as {\it Chandra\/}
and XMM-Newton.  Firstly, there is observational evidence that in
supernova remnants the post-shock electrons are cooler than the ions
\cite{hwang-snr, rakowski-snr}.  Secondly, the simple Rankine-Hugoniot
equations do not take into account the fact
that the bow shock around a lobe is oblique away from its head, with a
consequent change in the jump conditions and the emissivity contrast
\cite{williams}.  However,  in Sect.~\ref{sec:pressure},
we describe how the above equations can also
be applied to an overpressured lobe in the inner structure of the
low-power radio galaxy Cen~A.  The closer such a structure is to
a spherical expansion, the more normal the shock will be everywhere
and the better the applicability of the above equations.

\subsection{Low-power FRI jets}
\label{sec:fri}

\begin{figure}[t]
\centering
\includegraphics[height=9.0cm]{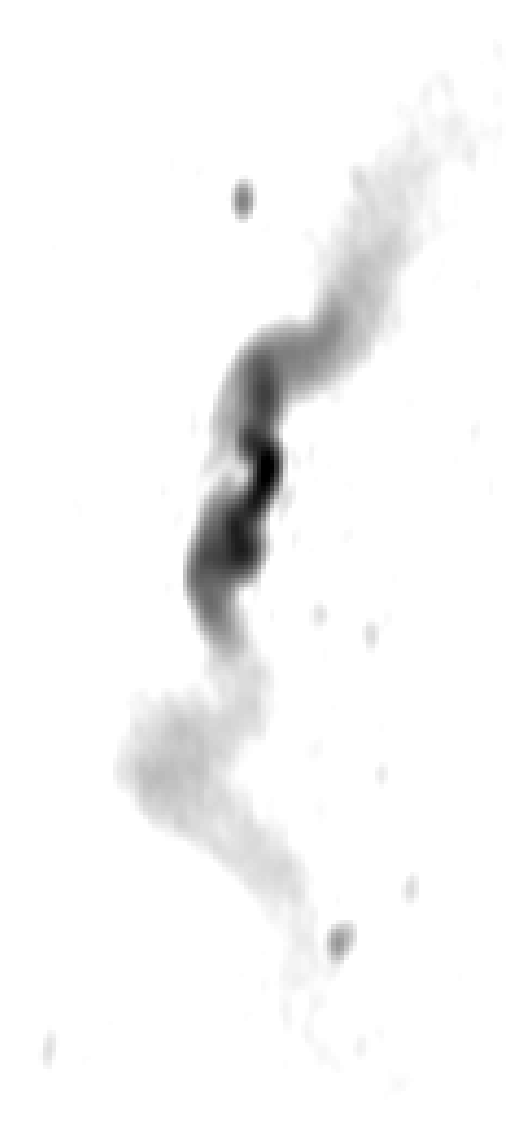}
\includegraphics[height=9.0cm]{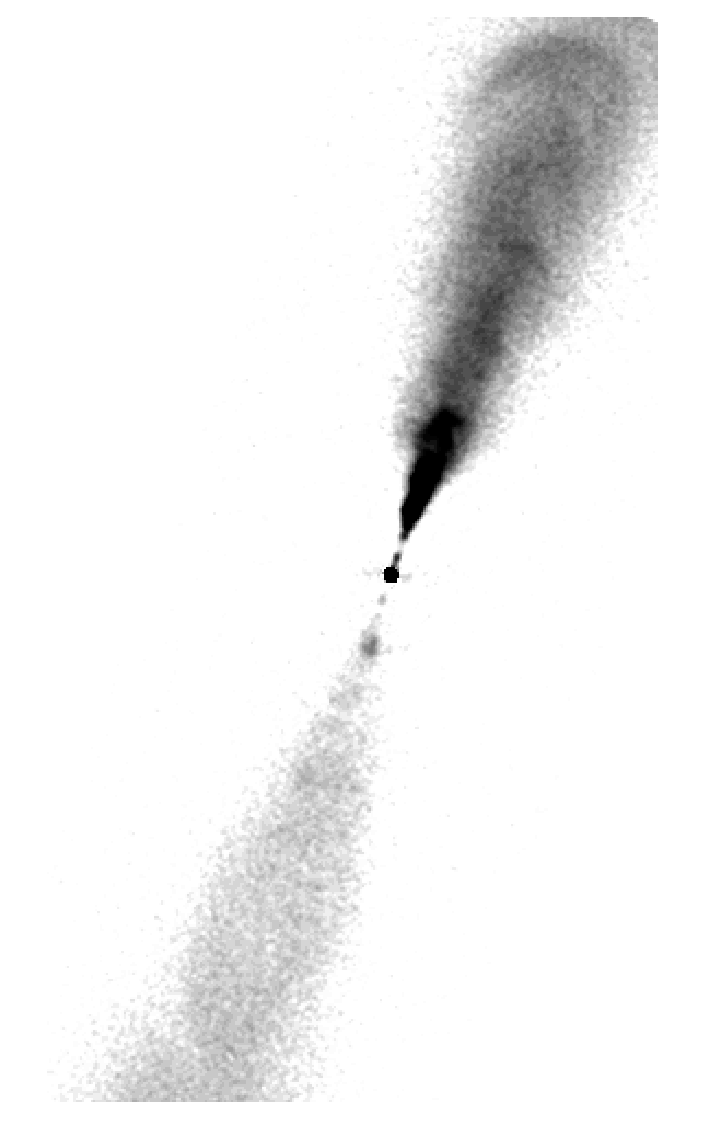}
\caption{Radio images of
an example low-power FRI source,
the radio galaxy 3C~31. 
The left image shows the large-scale structure.
The map \cite{leahy}
has a resolution of $29 \times 52$ arcsec and is from 
a 608~MHz observations using the
Westerbork Synthesis Radio Telescope.
The angular length is about 40 arcmin which,
at the galaxy's redshift of $z=0.0169$, corresponds to about 830 kpc
($H_o = 70$~km s$^{-1}$ Mpc$^{-1}$). The right image shows
an 8.4-GHz VLA map of the the inner $\sim$ 18 kpc where the jets
are relatively straight.
The resolution is 0.25 arcsec, and the data are from
\cite{laingbridle-31data}
}
\label{fig:radio31}   
\end{figure}

The appearance of low-power jets, in FRI radio galaxies, is quite
different from that of high-power jets: compare
Fig.~\ref{fig:radio220} with Fig.~\ref{fig:radio31}. Whereas the
high-power jets in FRII radio sources are generally weak features
within the well-defined, and much brighter radio lobes, low-power jets
are usually of high contrast against the inner radio
structures. Low-power jets are usually brightest near the active
galactic nucleus, and then fade gradually in brightness at larger
distances from the core, although this pattern is often interrupted by
patterns of bright knots in all wavebands.  FRI jets show a range of
morphologies, from almost straight and symmetrical two-sided jets to
the bent ``head-tail'' sources that are characteristic of low-power
extended radio sources in clusters.

While it is believed that the same basic mechanism is responsible for
the generation of low-power and high-power jets, it appears that
low-power jets dissipate much of their energy without developing a
well-defined beam head. This is often taken to indicate that low-power
jets are in good contact with the external medium in which they are
embedded, so that they share momentum and energy with entrained
material as they propagate. The strong velocity shear between a
fast-moving (originally relativistic, based on VLBI observations of
the cores of low-power radio sources \cite{giovannini}) jet flow and
the almost stationary external medium must generate instabilities at
the interface \cite{birkinshaw-kh}, and drive the flow into a
turbulent state. The physics of the resulting flow is far from clear,
although with simplifying assumptions \cite{bick-1} it has been shown
that subsonic and transonic turbulent jet flows can reproduce the
general trend of radio brightness and jet width for plausible
confining atmospheres. More recently, model fits \cite{hard-3c31}
using (\ref{eq:betasb}) and (\ref{eq:betaspecnorm}) to X-ray
measurements of the thermal atmosphere of 3C~31, coupled with results
from high-dynamic range radio mapping \cite{laingbridle-31data}
(Fig.~\ref{fig:radio31}), have led to a self-consistent model of the
flow.

The kinematic model of 3C~31 \cite{laing-bridle3c31} involves three
distinct regions in the jet: an inner $\sim 1$~kpc long section of
narrow jet where the flow is fast (relativistic) and the opening angle
of the jet is small, a flaring region of $\sim 2$~kpc during which the
jet broadens rapidly and brightens in the radio, and an outer region
in which the jet expands steadily with a smaller opening angle than in
the flaring region.  Here the jet decelerates steadily from a moderate
Mach number as it picks up mass from the external medium or from the
mass lost by stars embedded in the jet \cite{komissarov}.  Buoyancy
forces are important for much of the outer flow, and the jet becomes
sensitive to local changes in the density of the intracluster medium,
and hence is liable to deflect from straight-line motion.

While this model provides a good description of 3C~31, which displays
a smoothly-varying radio brightness profile and so is amenable to
simple fitting for flow speed, and is based simply on the basic
conservation laws for mass, momentum, and energy \cite{bick-n315}, it
represents only the overall deceleration of the jet from
entrainment. This leaves unanswered important questions about the
origins of the bright knots which appear in many jets, and which are
interpreted as the sites of strong shocks (which are relativistic,
based on proper motion studies), and about the detailed physics
responsible for mass entrainment and the deceleration of the jet.

While the knots in jets are sites where intense acceleration of
particles to highly relativistic energies can take place, it is clear
from the diffuse X-ray and other emission between the knots that there
is continuing particle acceleration even in the inter-knot
regions. The structures of the acceleration regions can only be
studied in detail in the closest radio galaxies, such as Cen~A
\cite{hard-cena} and M~87 \cite{m87-hstpolarim}. A number of features
of the knots have become apparent through detailed multiwavelength
studies, of which the most important are the presence of high-speed
motions (apparent speeds greater than the speed of light having been
noted in M~87), and the distinct locations for the acceleration of
particles radiating in different wavebands.

\section{The interpretation of multiwavelength data in terms of
physical parameters}
\label{sec:parameters}

\subsection{Minimum energy and magnetic field strength}
\label{sec:bmin}

The evidence for magnetic fields in the jets and lobes of active
galaxies (Sect.~\ref{sec:bfield}) confirms the synchrotron origin of
the radio emission, which is therefore an inseparable function of the
magnetic-field strength and electron spectrum.  To progress further
it is usual to assume that the source is radiating such that
its combined energy in particles and magnetic field is a minimum
\cite{burbidge}.  In this situation the energy in the magnetic field
is $\sim 3/4$ of the energy in the particles, and so this is similar
to the condition in which the two are equal and the source is in
``equipartition''.

We can calculate the minimum-energy magnetic field for a power-law
spectrum using the equations in Sect.~\ref{sec:synchrotron}, and for
more complicated spectra results can be obtained via numerical
integration.  The total energy in electrons

\begin{equation}
U_{\rm e} = \kappa\, m_{\rm e}c^2\, \int^{\gamma_{\rm max}}_{\gamma_{\rm min}} \gamma
\gamma^{-p} d\gamma = \kappa\, m_{\rm e}c^2\,  \int^{\gamma_{\rm max}}_{\gamma_{\rm
min}} \gamma^{-2\alpha} d\gamma
\label{eq:we1}
\end{equation}

\noindent
From (\ref{eq:synlumpowerlaw}), the synchrotron luminosity per unit frequency
is

\begin{equation}
L_\nu \propto \kappa \nu^{-\alpha} B^{\alpha + 1} 
\label{eq:we2}
\end{equation}

\noindent
and we can eliminate $\kappa$ from (\ref{eq:we1}) and (\ref{eq:we2})
to give (for $\alpha \neq 0.5$)

\begin{equation}
U_{\rm e} \propto L_\nu \nu^{\alpha} B^{-(\alpha + 1)}
\left(
\gamma_{\rm max}^{1 - 2\alpha} - \gamma_{\rm min}^{1 - 2\alpha}
\right) \left({1 - 2\alpha}\right)^{-1}
\label{eq:we3}
\end{equation}

If $K$ is the ratio of energy in heavy non-radiating particles to
that in electrons, and $\eta$ is the filling factor in the emission
region of volume $V$, the energy density in particles can be written
as

\begin{equation}
u_{\rm pcl} = C_1 B^{-(\alpha + 1)} {(1 + K)\over \eta V} L_\nu \nu^{\alpha} 
\left(
\gamma_{\rm max}^{1 - 2\alpha} - \gamma_{\rm min}^{1 -
2\alpha}
\right) (1 - 2\alpha)^{-1}
\label{eq:we4}
\end{equation}

\noindent
where the constant of proportionality, $C_1$, is determined
from synchrotron theory and can be expressed explicitly using the
equations in Sect.~\ref{sec:synchrotron}.  The energy density in the magnetic field can
be written as

\begin{equation}
u_{\rm B} = C_2 B^2
\label{eq:we5}
\end{equation}

\noindent
where the value of the constant $C_2$ depends on the system of units.
The value of $B$ which minimizes the sum of the two energy densities,
$B_{\rm me}$, is then given by

\begin{equation}
B_{\rm me} = \left[{(\alpha + 1)  C_1 \over 2 C_2} {(1 + K)\over \eta V} L_\nu
\nu^{\alpha}
 {\left(
\gamma_{\rm max}^{1 - 2\alpha} - \gamma_{\rm min}^{1 -
2\alpha}
\right) \over (1 - 2\alpha)}\right]^{1/(\alpha + 3)}
\label{eq:we6}
\end{equation}

\noindent
Since luminosity
density and flux density, $S_\nu$, are related by

\begin{equation}
L_\nu = (1+z)^{\alpha -1}\, S_\nu\, 4 \pi D_{\rm L}^2
\label{eq:we7}
\end{equation}

\noindent
where $D_{\rm L}$ is the luminosity density, and the volume of a radio
source can be specified \cite{miley} in terms of its angular component
sizes, $\theta_{\rm x}$, $\theta_{\rm y}$ and path length through the
source, $d$, as

\begin{equation}
V = \theta_{\rm x}\theta_{\rm y}\, d\, D_{\rm L}^2 / (1 + z)^4
\label{eq:we8}
\end{equation}

\noindent
we find

\begin{equation}
B_{\rm me} = \left[{(\alpha + 1) C_1 \over 2 C_2} 
{(1 + K)\over \eta\, \theta_{\rm x}\theta_{\rm y} d} 4\pi\, S_\nu\,
\nu^{\alpha}\, (1+z)^{3 + \alpha}
 {\left(
\gamma_{\rm max}^{1 - 2\alpha} - \gamma_{\rm min}^{1 - 2\alpha}
\right)\over ({1 -2\alpha})}\right]^{1/(\alpha + 3)}
\label{eq:we9}
\end{equation}

\noindent
At minimum energy, the total energy density is $u_{\rm me} = u_{\rm
B}(\alpha+3)/(\alpha + 1) = u_{\rm pcl}(\alpha + 3)/2$.  Any change in
the ratio of $u_{\rm B}$ to $u_{\rm pcl}$ increases the total energy and
pressure of the source.  Thus if, as seen in the southwest radio lobe
of Cen~A (Sect.~\ref{sec:pressure}), the minimum pressure is below
that of the external medium, expansion can immediately be inferred
if lobes are to stay inflated for an appreciable time.
Using minimum energy, the magnetic fields in radio sources are
generally estimated at 1-10 nT (10-100 $\mu$Gauss).

The dependence of $B_{\rm me}$ on the lower and upper
cut-off energies of the electron spectrum are weak, and since $\alpha$
is expected to be $\sim 0.6$ from shock acceleration
\cite{achterberg}, or larger where energy losses have steepened the
spectrum, it is the lower energy which has most effect.  There is a
stronger dependence of $B_{\rm me}$ on $(1 + K)$, the ratio of energy
in all particles to that in radiating electrons, and $\eta$.  It is
usual to adopt unity for both of these factors, which gives a true
minimum value for the energy and pressure [$(\gamma - 1)$ times the
energy density, where in this context $\gamma$ is the ratio of
specific heats, which is 4/3 for a relativistic gas].  Indirect arguments
are then applied to estimate if $(1 + K)$ should be larger than unity
or $\eta$ less than unity for a particular source.

In the case of SSC, $L_\nu$ must be converted into a spectral number
of photons per unit volume, $n_\nu$ [i.e., $n(\nu_o)$ in the equations
of Sect.~\ref{sec:compton}].
For a uniform spherical source, of radius $r$ and volume $V$, the
average value for $n_\nu$ is given by

\begin{equation}
n_\nu = {3\, r\, L_\nu \over 4\, c \, h\nu V}
\label{eq:nusyn}
\end{equation}

It is instructive to consider the effects of relativistic beaming
(Sect.~\ref{sec:doppler}).  For the case of a spherical blob,
(\ref{eq:we7}) is now the equation for $L_\nu \delta^{(3 + \alpha)}$,
and the total synchrotron luminosity in the source frame is reduced by
a factor of $\delta^4$, with $u_{\rm B}$ and $\kappa$, the
normalization of the electron spectrum, both reduced by $\delta^2$.
SSC in the source frame is reduced by a factor of $\delta^6$, and
thus sources which are more beamed have lower SSC to synchrotron flux
ratios.  [The factors of $\delta^2$ become
$\delta^{2(2+\alpha)/(3+\alpha)}$ for the jet case where $\delta^{(2 +
\alpha)}$ should be used.]

The minimum-energy assumption can be tested by combining measurements
of synchrotron and inverse-Compton emission from the same electron
population.  If the inverse Compton process is responsible for most of
the higher-energy radiation that is measured, and the properties of
the photon field are well known, the electron population is probed
directly (Sect.~\ref{sec:compton}).  In combination with measurements
of radio synchrotron emission from the same electrons, both the
electron density and magnetic field strength can then be estimated,
and the minimum-energy assumption can be tested. Since the application
of this test requires that the volume and
any bulk motion of the emitting plasma be known, the
best locations for testing equipartition are the radio hotspots, which
are relatively bright and compact, and are thought to arise from
sub-relativistic flows at jet termination (but see
\cite{georg-dechotspot}), and old radio lobes that should be in rough
overall pressure balance with the external medium. There is no reason
to expect dynamical structures to be at minimum energy.

{\it Chandra\/} has allowed such tests to be made on
a significant number of lobes and hotspots, with results
generally finding magnetic field strengths within a
factor of a few of their minimum energy (equipartition) values
for $(1 + K) = \eta = 1$  (e.g.,
\cite{brun-3c207, hard-hotspots}).  A study of $\sim 40$ hotspot
X-ray detections concludes that the most luminous hotspots tend to be
in good agreement with minimum-energy magnetic fields, whereas
in less-luminous sources the interpretation is complicated by an
additional synchrotron component of X-ray emission \cite{hard-hotspots2}.

\subsection{Jet composition}
\label{sec:composition}

Much of the detailed physics of jets depends on what they contain. 
While it is clear from their polarized radiation that jets
contain fast charged particles and magnetic fields, it is less clear
whether these are primary energy-carrying constituents of the jet or
secondary. A number of different possibilities for the energy carriers
have been suggested. The most popular are that the jets
are primarily composed of electrons and positrons, or of electrons and
protons, although electromagnetic (Poynting flux) jets \cite{rees71}
and proton-dominated jets \cite{mannheim,protheroe} have been
discussed. 

Since jets are presumed to obtain much of their energy from the
infall of matter into a supermassive black hole, it is natural to
suppose that electromagnetic radiation would carry much of the 
energy from the system on the smallest scales, since a plausible
mechanism for the extraction of energy from the system is the twisting
of magnetic field linked to the accretion disk. \cite{lovelace}
describe the processes that might launch such a jet, but fast
interactions with the plasma environment and efficient particle
acceleration are expected quickly to load the field with matter. In
the resulting magnetohydrodynamic flow much of the momentum
would be carried by particles, although Poynting flux
may carry a significant fraction of the total energy \cite{appl}. 

An electron-positron pair plasma is a natural consequence of the high
energy density near the centres of active galaxies, and hence it might
be expected that electrons and positrons would be an important, and
perhaps dominant, component of the jet outflow. Acceleration of such a
plasma by the strong radiation field near the active nucleus is
certainly possible \cite{odell,renaud}, although radiation drag is an
important limitation on the speeds to which the flow can be
accelerated if only electron-positron pairs are present \cite{sikora}.
The importance of electron-positron pairs can be established by
assuming that they are the major constituents of the flow, and then
testing that the kinetic energy that they carry down the jet is
comparable with the radiated power. 
While an electron-positron jet with arbitrary distribution of Lorentz
factors, $N(\gamma)$, can satisfy this condition, an electron-proton jet
should not extend to Lorentz factors $\gamma_{\rm min} < 100$, to
avoid a high order of inefficiency and depolarization by Faraday
rotation (Sect.~\ref{sec:bfield}).
Another test, based on the assumption that
the hotspots at jet termination in the FRII radio galaxy Cygnus~A
are reverse shocks in the jet fluid \cite{kino}, suggests that the jet
material is an electron-positron fluid at that point.

Further information supporting the interpretation of jets as
electron-positron plasmas at their point of injection has been
obtained by VLBI polarization studies \cite{wardle98}.  Here the
detection of circular polarization, interpreted as arising from mode
conversion, strongly suggests that, on the pc
scale, jets have low $\gamma_{\rm min}$ ($< 20$) and hence
are electron-positron plasmas  (Sec.~\ref{sec:bfield}), although \cite{ruskowski}
find that other possibilities (notably that the jet is composed of
electrons and protons) are not ruled out since other mechanisms exist
for the generation of circular polarization.

There are also grounds for expecting high-power jets to contain a
considerable fraction of their mass in protons. Protons outnumber
electrons by~100 to~1 in Galactic cosmic rays at energies at which
they leak out of the galaxy before significant energy loss
\cite{wolfendale}.  Some jet acceleration mechanisms can accelerate
heavy components of the jets as much as, or more than, the lighter
components, and, even if the jet is initially light, interactions with
the external medium are expected to load it with protons
(Sec.~\ref{sec:fri}). However, there is little direct evidence for the
proton contents of jets. Polarization studies do not find evidence of
embedded thermal material, and protons are relatively inefficient
radiators so the synchrotron emission of relativistic protons is
probably not detectable (but see \cite{aharonian}). It does seem to be
clear that jets are of low density relative to the external medium, on
the basis of numerical simulations, since the structures of dense jets
are unlike those observed, but there could still be appreciable mass
in the flow.

Support for the presence of protons is also found in models for the
broad-band spectral energy distributions of emission from some sub-parsec
scale quasar jets, where a significant proton
contribution is required to boost the kinetic power sufficiently to
match the total radiated power \cite{tav-qsocoresax}.  Some
radio lobes in low-power radio galaxies, if assumed to be radiating at
minimum energy (Sect.~\ref{sec:bmin}), would collapse because the
X-ray-emitting medium has a higher pressure \cite{crost-66b}. Although
there are several ways of boosting the internal pressure in such a
situation, magnetic dominance would make the sources unusual, electron
dominance can be ruled out by constraints on inverse-Compton
scattering of the CMB, and non-relativistic protons are disfavoured on
grounds of Faraday rotation (Sect.~\ref{sec:bfield}), leaving a
relativistic proton component most likely.
In contrast, there are other cases where the radio lobes would be
overpressured with respect to the ambient X-ray-emitting medium if a
relativistic proton contribution were included \cite{hard-hotspots}.
Although this in itself is not a major difficulty, since there are
open issues concerning lobe expansion (Sect.~\ref{sec:pressure}),
these are lobes for which the magnetic field can be measured
(Sect.~\ref{sec:bmin}) and the sources are close to minimum
energy when only the radiating electrons are considered. 

It has been known for some time that the minimum pressure in low-power
jets (calculated assuming an electron-positron plasma) is typically
below that of the external X-ray-emitting medium, e.g.,
\cite{feretti-b2rosat, killeen-n1399conf, morg-fripressure, 
worr-b2atm}.  However, this cannot be used simply to infer that the
jets are launched as an electron-proton plasma to give them the extra
required pressure.  Low-power jets are believed to slow down
to sub-relativistic speeds via entrainment of thermal
material (Sect.~\ref{sec:fri}), and even the most detailed
hydrodynamical modeling, such as that which has been applied to 3C\,31
by \cite{laing-bridle3c31}, does not decide the issue of primary jet
content.

On balance it appears that the dominant energy-carrying constituents
of jets are electrons and positrons, but that relativistic protons may
also be important, particularly in momentum transport. However, the
situation remains unclear, and there are degeneracies between the
measurement of jet speed and jet composition which render this
conclusion tentative.

\subsection{Jet speed}
\label{sec:speed}

The resolved X-ray jets (excluding hotspots) that {\it Chandra\/}
detects in powerful radio sources are mostly in quasars, e.g.,
\cite{chartas-pks0637, marsh-chandra3c273, samb-3c273, samb-qso17,
schwar-pks0637, siem-1127, siem-1508z=4.3, siem-0738}, with the bright
radio sources Pictor~A and Cygnus~A \cite{wilson-pica, wilson-cygajet}
being exceptions.  The quasar X-ray jet emission is one-sided, on the
same side as the brighter radio jet, implying that relativistic
beaming is important.  Two-sided X-ray emission, such as that in the
quasar 3C~9 \cite{fab-3c9} most likely does not imply the presence of
counter-jet emission, but rather the presence of the more isotropic
lobe, hotspot, or cluster-related emission expected at some level in
all sources and detected in many, e.g., \cite{hard-hotspots,
worr-c220.1}.

Currently there are several tens of quasar X-ray jet detections.  They
have mostly been found through targeted programs to observe bright,
prominent, one-sided radio jets.  In most cases there has been no
pre-existing reported optical jet detection, but there has been
reasonable success from follow-up work.  The level of many
such detections lies below an interpolation between the radio and X-ray
spectra, suggesting that
synchrotron emission from a single power-law
distribution of electrons is not responsible for all the emission,
e.g., \cite{samb-qso17, schwar-pks0637}, although it has been
pointed out that since high-energy electrons lose energy less
efficiently via inverse Compton scattering in the Klein-Nishina
regime, the electron spectrum should harden at high energies and in
some cases may produce a synchrotron spectrum that can match
observations
\cite{dermer-atoyan}.

In order to avoid the total energy in particles and magnetic field
being orders of magnitude above its minimum value, as would arise from
a simple SSC explanation, the most widely favored model for the X-ray
emission in these cases is that the
X-rays are produced by inverse Compton scattering of CMB photons by
the electrons in a fast jet (Sect.~\ref{sec:doppler}) that sees
boosted CMB radiation and emits beamed X-rays in the observer's frame
\cite{cel-pks0637, tav-pks0637}.  The model can produce sufficient
X-rays with the jet at minimum energy, but only if the bulk motion is
highly relativistic (bulk Lorentz factor, $\Gamma \approx 5-20$) and
the jet at a small angle to the line of sight.  Although such a speed
and angle are supported on the small scale by VLBI measurements, at
least for the source which has guided this work, PKS 0637-752
\cite{lovell-pks0637, schwar-pks0637}, the jet must remain highly
relativistic hundreds of kpc from the core (after projection is taken
into account) for the X-rays to be produced by this mechanism.  This
conclusion, based on multiwavelength data, has been something of a
surprise, since earlier statistical studies of the structures of
powerful radio sources suggested that jet velocities average only
about $0.7c$ at distances of tens of kpc from the core
\cite{hard-zlt0.4frII, war-jetspeed}.

A difficulty with the interpretation of the X-rays from radio jets as
due to inverse-Compton scattering of CMB photons by a fast jet is that
the observed gradients in X-ray surface brightness at the edge of the
knots are sharp \cite{tavech-clumps}. Since the X-rays are generated
from low-energy electrons (as the CMB photons are boosted in the electron
rest frame), the lifetimes of the electrons are long and it is
difficult to see how X-ray knots in high-power jets can have a steeper
gradient (as in PKS 0637-752, \cite{schwar-pks0637}) than radio knots,
which are produced by electrons of similar or higher energy. Although
\cite{tavech-clumps} suggest that the effects of strong clumping in
the jets may resolve this issue, a fast jet and the proposed mechanism
is then no longer required, since such clumping would increase the SSC
yield for a slow jet at minimum energy \cite{schwar-pks0637}.

\subsection{Particle acceleration in jets}
\label{sec:acceleration}

The electrons that emit synchrotron radiation in the radio, optical,
and X-ray wavebands are all ultrarelativistic (with Lorentz factors
$> 10^3$). The existence of large numbers of such relativistic
electrons depends on their
acceleration to high energies {\it locally} within the jet, or
hot-spot, since their lifetimes against synchrotron losses
(Sec.~\ref{sec:synchrotron}) are usually less than the minimum
transport times from the active nuclei.
(This may not be the case if proton synchrotron radiation is important
\cite{aharonian}.)
Particle acceleration is 
generally discussed for the cases of a particle interacting with a
distributed population of plasma waves or magnetohydrodynamic
turbulence, or shock acceleration. Reviews of these processes may be
found in \cite{blandford-eichler, drury, eilek} and elsewhere.

Resolved X-ray jets in active galaxies with {\it low\/} radio power
are detected with {\it Chandra\/} in sources covering the whole range
of orientation suggested by unified schemes, suggesting that beaming
is less important than in their more powerful counterparts.  The
several tens of detected sources range from beamed jets in BL Lac
objects \cite{birk-pks0521, pesce-3c371} to two-sided jets in radio
galaxies \cite{chi-n4261, hard-cena}, with most X-ray jets
corresponding to the brighter radio jet, e.g., \cite{hard-66b,
hard-3c31, harris-3c129, harris-m84, marsh-hetgsm87, worr-fr1s,
worr-n315}.  Several of the observations have been targeted at sources
already known to have optical jets, from ground-based work or HST.
However, it's easier \cite{worr-fr1s} to detect X-ray jets in modest
{\it Chandra\/} exposures than to detect optical jets in HST snapshot
surveys, because there is generally better contrast with galaxy
emission in the X-ray band than in the optical.

Inverse Compton models for any reasonable photon field suggest an
uncomfortably large departure from a minimum-energy magnetic field in
most low-power X-ray jets, e.g., \cite{hard-66b}.  Synchrotron
emission from a single electron population, usually with a broken
power law, is the model of choice to fit the radio, optical, and
X-ray flux densities and the relatively steep X-ray spectra, e.g.,
\cite{boer-m87xmm, hard-66b}.  X-ray synchrotron emission requires
TeV-energy electrons which lose energy so fast that they must be
accelerated {\it in situ}.

The above arguments, applied to the bright northeast jet of the
nearest radio galaxy, Cen~A, find in favor of X-ray synchrotron
emission \cite{kraft-cenajet}, and the proximity of Cen A allows its
acceleration sites to be probed in the greatest possible detail.
Unfortunately the dramatic dust lane spanning the galaxy masks any
optical jet emission.  Proper motion of order 0.5c, observed both in
the diffuse emission and some knots of Cen~A's radio jet, is
indicative of bulk motion rather than pattern speed
\cite{hard-cena}. Since Cen~A has a strong jet to counter-jet
asymmetry, ({\ref{eq:jetcounterjet}) then suggests that the jet is at
a small angle to the line of sight.  Since this contradicts the
evidence based on parsec-scale properties, and other considerations,
that the jet is at about 50 degrees to the line of sight, Cen~A
appears to be a case where intrinsic effects render
({\ref{eq:jetcounterjet}) inapplicable.  Some of the bright X-ray
knots have only weak radio emission with no indication of proper
motion, but with the radio emission brightening down the jet in the
direction away from the nucleus.  While the radio association confirms
that these X-ray knots are indeed jet related, the emission profiles
are not what are expected from a simple toy model where the electrons
are accelerated and then advect down the jet, with the X-ray emitting
electrons losing energy faster than the radio-emitting electrons.
Instead, it is proposed \cite{hard-cena} that there are obstacles in
the jet (gas clouds or high-mass-loss stars).  Both radio and
X-ray-emitting electrons are accelerated in the standing shock of this
obstacle, and a wake downstream causes further acceleration of the
low-energy, radio-emitting, electrons.  The resulting radio-X-ray
offsets, averaged over several knots, could give the radio-X-ray
offsets seen in more distant jets, e.g., \cite{hard-66b}.

The X-ray and radio emission in hot-spots are also offset in some
cases \cite{hard-hotspots}, presumably also because of the
different locations of acceleration of the particles at the different
energies probed in these wavebands, if the emission is all
synchrotron. However, it is expected that much of the X-ray emission
should be of inverse-Compton origin, and then the offsets become
harder to understand, although \cite{georg-dechotspot} have described a
model in which decelerations of the jet plasma near a hot-spot can
generate X-ray enhancements and small offsets.

Optical polarization might help us to learn more about the
acceleration processes in low-power jets.  In M~87 there is evidence
for strong shock acceleration at the base of bright emitting regions,
in compressed transverse magnetic fields \cite{m87-hstpolarim}.  A
knot in the jet of M 87 has been observed to vary in the X-ray and
optical on the timescale of months, consistent with shock
acceleration, expansion, and energy losses \cite{harris-hstvarm87,
perl-m87hstvar}.

\subsection{Pressure and confinement}
\label{sec:pressure}

It is interesting to compare the minimum pressure
(Sect.~\ref{sec:bmin}) in radio lobes with that of the external
X-ray-emitting environment.  Over-pressure in the lobe would imply an
expansion, which may be supersonic and should involve significant
heating of the external gas.  Under-pressure suggests either that the
lobe is undergoing collapse (this should be rare given their
prevalence) or that there is an additional component of pressure that
may be in the form of relativistic protons.  FRII sources are
generally at high redshift where observations lack sufficient
sensitivity and resolution to draw strong conclusions.  However,
notwithstanding the fact that there is no reason to expect dynamical
structures to be at minimum energy, where tests are possible it tends
to be confirmed (Sect.~\ref{sec:bmin}), and it appears that with this
assumption rough pressure balance prevails, e.g., \cite{belsole,
croston2, hard-hotspots}.

The situation in low-power radio galaxies is more complicated, because
the medium plays an important role in the deceleration of the jets,
such that they share momentum and energy with entrained material.
However, these sources have the advantage of being closer, and
Sect.~{\ref{sec:fri} describes how the pressure profile predicted by
the mass-entrainment model for deceleration of the jet in 3C~31 gives
an excellent match to observations.  The outer radio structures of
FRIs may sometimes be buoyant, e.g., {\cite{worr-n346}, and in other
cases show evidence of having done significant work on the gas in
pushing it aside \cite{boer-n1275rosat, mcnam-hydraA}, or
responsibility for lifting gas in hot bubbles, e.g., \cite{churazov-m87},
expected to result in eventual heating,
e.g., \cite{quilis-bubbles}.
Such heating  would
help to explain the weakness or absence of lines from gas cooling
below 1~keV in the densest central regions of galaxy and cluster
atmospheres, e.g., \cite{peter-a1835cool}.
A statistical study shows
that atmospheres containing radio sources tend to be hotter
than those without \cite{crost-66b}.

One place where heating is definitely expected is from gas crossing the
supersonically advancing bow shock of an expanding lobe
(Sec.~\ref{sec:frii}).  It is possible to interpret X-ray cavities
coincident with the inner parts of the radio lobes of Cygnus~A as due
to an emissivity contrast between bow-shock heated gas outside the
lobes and the more easily detected ambient cluster medium
\cite{carilli-cyga}, although the parameters of the shock are not
effectively constrained by the data.  More recent {\it Chandra\/}
observations of Cygnus~A find gas at the sides of the lobes to have
$kT \sim 6$~keV, slightly hotter than the value of 5~keV from ambient
medium at the same cluster radius, possibly indicating cooling after
bow-shock heating, but again the data do not usefully constrain model
parameters \cite{smith-cygacluster}.

\begin{figure}[t]
\centering
\includegraphics[height=10cm]{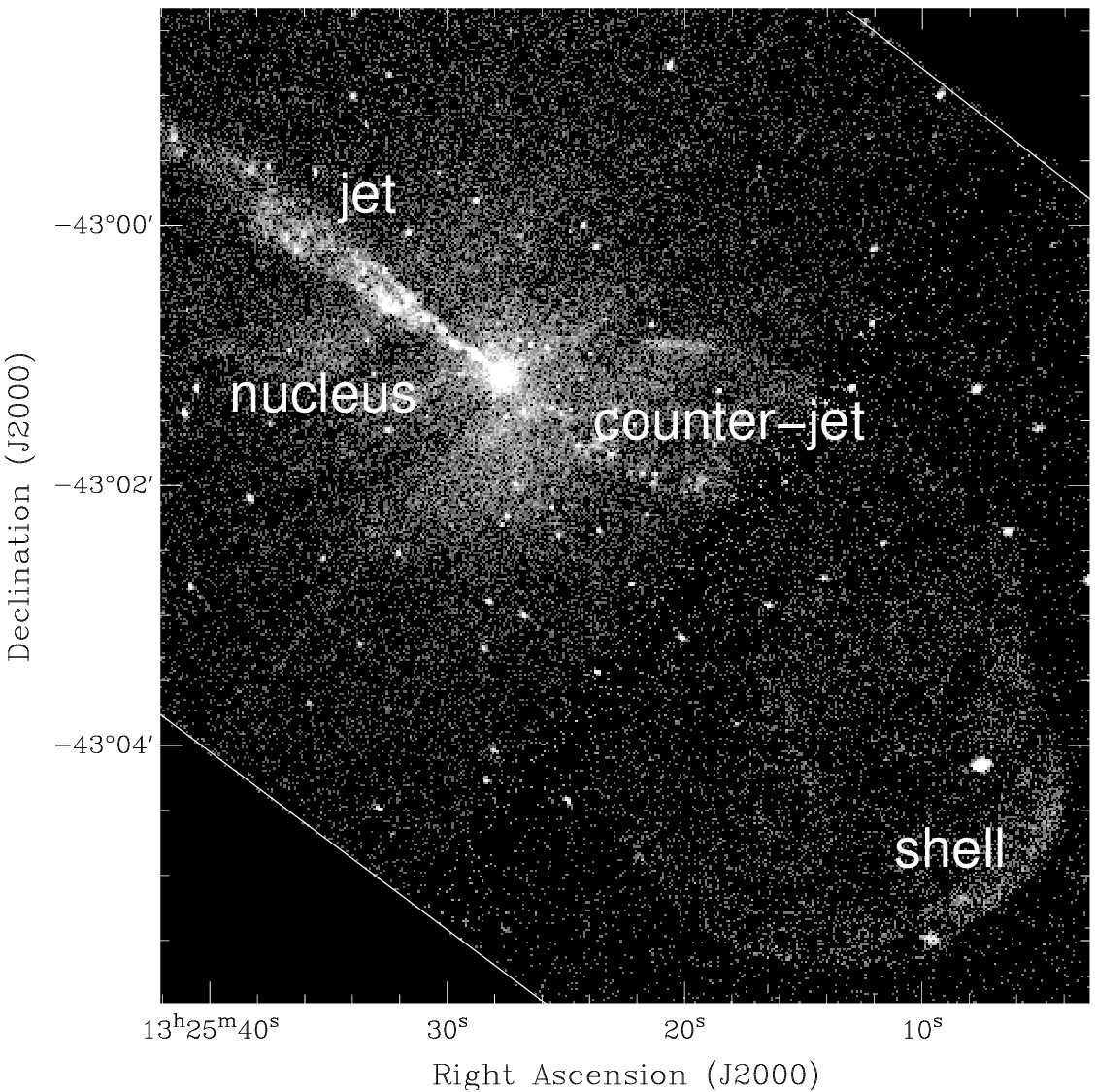}
\caption{0.5--5~keV X-ray image of Cen~A from a 50~ks {\it Chandra\/}
exposure using the ACIS-S instrument. Pixel size is 1~arcsec. Data are
exposure-corrected.
Many point sources associated with Cen~A's host galaxy NGC~5128 are
seen, together with a diffuse background of X-ray-emitting gas of
$kT \approx 0.3$ keV, and labelled X-ray structures that are in whole
or part related to radio structures (see the text and
\cite{evans-cena, hard-cena,
kraft-cenajet, kraft-cenalobe} for more details) 
}
\label{fig:cenaacis}  
\end{figure}

The first and best example of a shell of
heated gas which can reasonably be attributed to supersonic expansion
is not in an FRII radio source, but in Cen~A
(Fig.~\ref{fig:cenaacis}).  High-quality {\it Chandra\/} and XMM-Newton data
\cite{kraft-cenajet} provide the temperature and density
constraints needed to test the model and measure the supersonic
advance speed of the bow shock responsible for the heating.
This source is an excellent example where much of the theory
outlined earlier in this chapter can be applied.

Cen~A is our nearest radio galaxy, at a distance of 3.4~Mpc
\cite{israel-cena} so that
1~arcsec corresponds to $\sim 17$~pc.
The full extent of Cen~A's radio
emission covers several degrees on the sky \cite{junkes-cena}.
Within this lies a sub-galaxy-sized double-lobed inner structure
\cite{burns-cena} with a predominantly one-sided jet to the northeast
and a weak counter-jet to the southwest \cite{hard-cena}, embedded in a
radio lobe with pressure $1.4 \times 10^{-12}$~Pa or more, greater
than the pressure in the ambient ISM ($\sim 1.8 \times 10^{-13}$~Pa;
Table~\ref{tab:physical}), and so which should be surrounded by a
shock.  Around this southwest lobe there is a shell of X-ray emitting
gas which appears to have the geometry of the shocked ambient gas in
Fig.~\ref{fig:supersonic} \cite{kraft-cenalobe}.  Although the
capped lobe is around the weak counterjet, so it is not evident that
the lobe is being thrust forward supersonically with respect to the
external interstellar medium (ISM) by the momentum flux of an active
jet, 
the current high internal pressure in the radio lobe
ensures its strong expansion.

\begin{table*}[t]
  \begin{center}
  \caption{Physical parameters of the gas in various regions of Cen
  A}\medskip
  \label{tab:physical}
  \begin{tabular}{lcclc}
Structure & $kT$ & n$_{\rm p}$ & Pressure &
  0.4--2 keV  \\
        & (keV) & (m$^{-3}$) & (Pa)$\dag$ &
   relative \\
&& &&emissivity, $\cal E$ \\
    \hline\hline
ISM (measured) & 0.29 & 1700 & $1.8 \times 10^{-13} \ddag$ &
  1 \\
Behind bow shock (inferred) & 6.8 & 6530 & 
$2.1 \times 10^{-11}$  & 13\\
Shell (measured) & 2.9 & 20000 & $2.1 \times 10^{-11}$ &
  127\\
    \hline
$\dag$  1 Pascal = 10 dyn cm$^{-2}$ & \multicolumn4{c}{$\ddag$
  incorrectly
  reported in Table~5 of \cite{kraft-cenalobe}}\\
\multicolumn5{l}{ISM and Shell pressures are thermal only. Inferred
  pressure includes ram pressure} \\
\multicolumn5{l}{ of $\rho_1 v^2_{\rm adv} ({\cal M}^2 + 3)/ 4 {\cal M}^2$}. \\
\multicolumn5{l}{
The minimum-energy pressure in the radio lobe is $\sim
  1.4 \times 10^{-12}$ Pa.}\\
  \end{tabular}
  \end{center}
\end{table*}

The temperature, proton density and pressure of the ambient ISM and
the X-ray shell, taken from \cite{kraft-cenalobe} are given in
Table~\ref{tab:physical}.  The ambient medium is measured to have
$n_{\rm p}\sim 1.7 \times 10^3$ m$^{-3}$ and $kT = 0.29$ keV,
whereas the shell is ten times hotter, at $kT = 2.9$ keV, and
twelve times denser, with $n_{\rm p}\sim 2 \times 10^4$ m$^{-3}$.
From (\ref{eq:rhtemp}) and (\ref{eq:rhdens}), we see that temperature and density
measurements for both the ambient medium and the shocked gas directly
test shock heating, since only two of the four parameters are required
to measure the Mach number, and the other two test the model.  

The most straightforward application of the equations finds that the
densities and temperatures are not self-consistent.  The shell's
density and temperature are wrong for gas directly in contact with the
bow shock.  However, we can find a Mach number consistent with
shocking the gas to a temperature and density such that the combined
thermal and ram pressure is in pressure equilibrium with the thermal
pressure of the detected shell: ${\cal M} = 8.5, v_{\rm adv} \approx
2400$ km s$^{-1}$.  The post-shock temperature is $kT_2 \sim 6.8$~keV.
The 6.8~keV gas flows back from the shock, into the X-ray-detected
shell at 2.9~keV.  The characteristics of this undetected hotter gas
are given in Table~\ref{tab:physical}.  In this table we also quote
estimates of the relative X-ray emissivity (per unit volume) of gas in
the different structures over the 0.4-2 keV energy band, where the
{\it Chandra\/} response peaks and is relatively flat.  The gas
directly behind the bow shock has a predicted emissivity that is an
order of magnitude fainter than that in the shell, accounting for its
non-detection in our measurements.

The radiative timescale for material in the shell is $\sim 2
\times 10^9$
yrs (Fig.~\ref{fig:cooling}), which is large compared with
the lobe expansion time ($< 2.4 \times 10^6$
years), so the material in the shell behaves as an adiabatic
gas \cite{alexander}.
The shell is overpressured compared with the minimum-energy
pressure in the radio lobe (in magnetic field and radiating electrons)
by a factor of $\sim 10$.  If we assume minimum energy in the lobe,
despite it being a dynamical structure,
and that the shell has reached equilibrium [but note that
the sound-crossing time in the shell (thickness $\sim 0.3$
kpc, $c_{\rm s} \sim  9 \times 10^{-7}$ kpc yr$^{-1}$)
is about 15 per cent of the maximum time we estimate it has taken the
lobe to reach its current size], the shell's  overpressure relative to the
radio lobe could be balanced by the ram pressure from internal motions in
the lobe for a moderate relativistic proton loading.
Finally, the shell's kinetic energy is $\sim 5$ times its thermal
energy, and exceeds the thermal energy of the ISM within 15 kpc of the
centre of the galaxy.  As the shell dissipates, most of the kinetic
energy should ultimately be converted into heat and this will
have a major effect on Cen~A's ISM, providing distributed heating.

\subsection{Magnetic field structures}
\label{sec:bfield}

Jets contain significant magnetic fields.
The orientation of these fields is
displayed by the linear polarization that they show in their radio 
synchrotron emission, and for many years detailed maps of the radio
polarizations of jets have been used to infer the magnetic field
geometries both on kpc scales, e.g., for NGC~6251 \cite{perley-n6251} and
3C~31 \cite{fomalont-3c31}, and pc scales, e.g., for BL~Lac objects
\cite{cawthorne-gabuzda}. 

The fractional linear polarization expected from an optically-thin
syn\-chro\-tron-emitting plasma is

\begin{equation}
\pi_{\rm L} = {{\alpha + 1} \over {\alpha + {5 \over 3}}}
\label{eq:linpol}
\end{equation}

\noindent
where $\alpha$ is the radio spectral index, and it is assumed that
the frequency of observation satisfies $\gamma_{\rm
min}^2 \nu_{\rm g} \ll \nu \ll \gamma_{\rm max}^2 \nu_{\rm g}$, as
for (\ref{eq:synlumpowerlaw}).
The electric field of the emitted radiation is perpendicular
to the magnetic field within which the emitting electron population
lies, so that an observation of the plane of polarization of the radio
emission can be used to infer the geometry of the magnetic field
within the radio jet.

In mapping the direction of the projected magnetic field, account has
to be taken of the effect of Faraday rotation of the plane of
polarization by mixed thermal material and fields within and around
the radio jets. For a foreground plane slab of plasma, at a wavelength
of $\lambda$, the Faraday
rotation is an angle 

\begin{equation}
   \theta_{\rm F} = {\rm RM} \, \lambda^2
\label{eq:frangle}
\end{equation}

\noindent
where the rotation measure along path length $L$ is

\begin{equation}
  {\rm RM} = {1 \over c
} \int_0^L \, n_{\rm e,th}\, r_{\rm e} \, \nu_{\rm g} \, \cos
  \psi \, dz
\label{eq:rm}
\end{equation}

\noindent
where $n_{\rm e,
th}$ is the thermal electron density,
$r_{\rm e}$ is the classical electron radius, $\nu_{\rm g}$ is the non-relativistic
gyrofrequency, and $\psi$ is the angle of the magnetic field to the
line of sight.  In SI units,

\begin{equation}
  {\rm RM} = 8100 \int_0^L \, n_{\rm e,th} \, B_\parallel \, dz
  \quad{\rm rad~m}^{-2}
\label{eq:rmsi}
\end{equation}

\noindent
for a magnetic field with line-of-sight component $B_\parallel$ (in
Tesla), where $n_{\rm e,th}$ is in $\rm m^{-3}$ and the
thickness $L$  is in pc. If the thermal material is
mixed with the radio jet's plasma, an effect half this size is
expected. If the magnetic field and thermal material have a
complicated structure, different dependencies of $\theta_{\rm F}$ on
$\lambda$ are possible. Faraday rotation is measured by
multi-frequency radio mapping, and uncertainties in the intrinsic
plane of polarization of the radio emission are minimized by working
at high radio frequencies.

Since the presence of an appreciable density of thermal material
around the jets is revealed in X-rays (Sect.~\ref{sec:medium}), a
significant Faraday rotation is possible if this medium contains a
magnetic field, and some evidence for such Faraday rotation has been
seen, e.g., in Hercules~A \cite{hera-leahy}, with the brighter, and
presumably Lorentz-boosted, jet often showing a lower Faraday rotation
than the opposite jet, as might be expected from simple geometrical
considerations \cite{garrington, laing}.

A more important consequence of the significant linear polarizations
measured for radio jets, however, is that those jets cannot contain
large amounts of thermal material. If they did, then the change in
$\theta_{\rm F}$ between the front and back sides of the jets would
lead to linear polarizations far less than the observed values, which
are sometimes near the theoretical maximum given by
(\ref{eq:linpol}). The limits on Faraday depolarization thus lead us
to conclude that the X-rays from radio jets and lobes cannot come from
thermal material mixed with the jets.

Polarization mapping also reveals that the magnetic fields in radio
jets are relatively well ordered. In many jets of both high and low
radio power the magnetic fields are longitudinal for the first few kpc
of the length of the jet, then, in lower-power jets, the
magnetic field becomes perpendicular to the jet axis at larger
distances from the core, often after a significant change in the jet's
width \cite{bridle-perley}. Variations on this pattern are seen where bright
emission knots exist in the jet. These knots appear to be associated
with strong shocks, which compress the magnetic field strongly into a
transverse pattern at the upstream (core-side) of the knot, after
which the magnetic field pattern becomes complicated.  A particularly
good example  \cite{m87-hstpolarim} is M~87, where the use of HST to
map the optical polarization provides additional information, since
the lifetimes of the electrons emitting optical synchrotron emission
are far shorter than those of radio-emitting electrons.
Significant differences between the polarization structures seen
in the optical and radio in M~87 suggest that the sites of acceleration
are different for different electron energies, with the strongest
shocks, that provide acceleration to the highest energies, appearing
in the most central parts of the jet.

In the central parts of radio sources, VLBI observations have been
able to detect both linear and circular polarization. A distinct
difference is seen in the polarizations of the pc-scale (VLBI) jets of
BL~Lac objects and radio galaxies or quasars.  Generally, the magnetic
field orientation in the cores of the highest-power sources is along
the jet \cite{cawthorne-gabuzda}, but in lower-power sources, such as
BL~Lac objects, regions of significant transverse field are seen, and
change on short timescales \cite{gabuzda-cawthorne}. Again this is
interpreted as the effect of shock structures moving along the jets,
compressing the magnetic field as they pass, although helical
patterns in the flow may affect the apparent field pattern, as may the
effects of relativistic aberration. The lower linear polarizations
often seen in pc-scale jets may be due to depolarization caused by the
superposition of many small-scale structures, with different field
orientations, within the resolution of the observations, or perhaps to
significant structure in the nuclear gas near the jet.

Circular polarization is generally undetectable in large-scale jets,
but has been mapped in pc-scale jets using VLBI \cite{homan_wardle_1,
homan_wardle_2}, where it is interpreted as arising from mode
conversion --- the conversion of linear polarization to circular
polarization by Faraday rotation in the jet plasma, although this is
not certain \cite{ruskowski}.
If the X-ray emission
from powerful jets on the large scale is interpreted as
inverse-Compton scattering of CMB photons in a fast jet, then the
population of relativistic electrons has to extend down to low
energies (Lorentz factors $\gamma_{\rm min} = 10 - 20$,
\cite{cel-pks0637,tav-pks0637}).  On the other hand, the detection of
significant polarization in VLBI jets implies that there should
be few low-energy protons and electrons (which would cause excessive internal
Faraday rotation), and hence that the jets should be composed of an
electron-positron plasma.  However, the presence
of an electron-positron plasma in the VLBI jet does not necessarily
imply its presence on the largest scales, where the jet may have
become loaded with material gathered from the ambient medium
or the radio lobe.

\subsection{Cores: the inner jets}
\label{sec:cores}

Radio jets extend into the cores of active galaxies, where they are
faster and more compact.  Special-relativistic effects then
cause their brightness and variability time scales to be strong
functions of jet orientation.
As a result of synchrotron self-absorption,
it is important to use VLBI techniques at high radio or mm frequencies
to see close to the bases of jets.

The inner jets are difficult to distinguish at non-radio wavelengths
because of their closeness to the central engine, which is bright at
infrared to X-ray energies, and because of orientation-dependent
absorption in the optical to soft-X-ray bands from gas and dust
structures.  These gas and dust structures are only sometimes
detected, but their presence is often inferred so that models which
unify various classes of active galaxy might work,
e.g., \cite{barthel}.  They affect the central engine the most.  Thus,
the first problem in using multiwavelength data to gain additional
insight into radio jets on small scales is to separate out the jet
emission.  For no waveband does telescope spatial resolution match
that of radio VLBI, and so only the tool of spectral separation is
usually available.

In the most core-dominated radio sources, such as bright variable BL
Lac objects and quasars (often classed together as blazars), jet
emission appears dominant at all energies, sometimes up to TeV.  The
multi-wavelength spectral energy distributions and variability time
scales are used to probe the beaming parameters and the physical
properties of the emitting regions, e.g.,
\cite{ghisellini-model, krawczynski-mrk421tevvar, tagliaferri-var}.  
Correlated flares are sometimes
measured across wavebands, giving support to the present of a dominant
spatial region of emission, e.g., \cite{takahashi-mrk421var,
urry-2155var}, but otherwise uncertainties of size scales, geometries,
and parameters for the competing processes of energy loss and
acceleration, often force the adoption of oversimplified or
poorly-constrained models for individual jets.

In the quasar population in general, there is good evidence that
in the X-ray an inner jet dominates the emission from core-dominated
quasars, but not lobe-dominated quasars,
e.g., \cite{browne-murphy, kembhavi, worr-giommi, worr-c280, zamorani}.
Radio galaxies are of particular interest, because here emission from
the central engine is weak.  In these sources a correlation of
core soft-X-ray and radio emission, lost in lobe-dominated quasars,
re-appears \cite{canosa, fabbiano-3c, 
hard-worr, worr-rosfr1}, suggesting that jet-related X-ray emission is
again dominant.  In low-power radio galaxies, an optical core
is often seen with HST, and is interpreted as synchrotron emission from a
similar small-scale emitting region \cite{capetti-b2optcores,
chi-optcores, hard-worroptcore, verdoes-kleijn}.
However, at higher X-ray energies, a number of radio galaxies show
clear evidence of a hard, highly absorbed continuum,
sometimes accompanied by Fe-line emission, e.g., \cite{glioz-n4261core, ueno}.  Here
the decreasing jet output is leading to the central engine becoming
increasingly dominant.  Both X-ray
components can sometimes be distinguished in the same
spectrum, e.g., \cite{croston2, evans-cena}.

The richness of high-energy structure in larger-scale radio jets
has been revealed because it is well resolved with HST and {\it Chandra}.
Our knowledge of the inner jets is limited by the confusion of
components that has now been lifted for the jets discussed earlier
in this chapter.

\section{Conclusion}
\label{sec:conclusion}

The new results on radio jets which have resulted from complementary
X-ray and optical observations have brought some surprises.  Firstly,
synchrotron X-ray jets are common in low-power sources, which implies
that the intrinsic electron spectrum continues to TeV energies, and
requires substantial {\it in situ\/} particle acceleration.  Secondly,
the detection of many quasar X-ray jets, most commonly interpreted
as due to beamed CMB photons, would suggest that highly relativistic
bulk flows exist far from the cores, contradicting
earlier statistical studies of radio sources
\cite{hard-zlt0.4frII, war-jetspeed}.  Jet theory has had some
pleasing successes, such as the agreement of the X-ray pressure
profile with the prediction from a hydrodynamical model for 3C~31
\cite{laing-bridle3c31}.

There is still much observational work to be done.  Firstly, there is
considerable bias in the jets which have been observed in the X-ray,
and we need observations of unbiased samples over broader luminosity
and redshift ranges, together with observations of the X-ray-emitting
medium through which the jets propagate.
The measurements should elucidate the relative importance
of synchrotron emission,
inverse Compton scattering, and relativistic beaming,
refine our knowledge of the source energetics, and improve
constraints on jet composition and speed.
Secondly, we need more deep X-ray observations
(and refined theory) to understand jet-lobe/intercluster medium
interactions.  Finally, to study acceleration sites and processes,
deeper and more detailed multiwavelength mapping, spectroscopy, 
and temporal monitoring is required.  In
combination with multifrequency polarization measurements, such data
could map the spatial distributions and follow the acceleration of the
electrons responsible for the radiation in the radio to X-ray bands.

It should, however, be acknowledged that much basic physics of jets is
still not well understood, e.g., the origin of the magnetic field, the
method of jet production and collimation, and the effects of
turbulence.  Much theoretical work in these areas is necessary to
relate measurements to the processes occurring in the fascinating jet
flows that are observed.

\input{worrall-refs}

\end{document}

%% file: worrall-refs.tex
%
%

%
%